\documentclass[aps,pra,floatfix,twocolumn,tightenlines,amsmath,amssymb,nofootinbib]{revtex4-1}

\usepackage{graphicx}
\usepackage{amssymb}
\usepackage{color}
\usepackage{psfrag}
\usepackage{ifsym}
\usepackage{epstopdf}
\usepackage{dsfont}
\usepackage{amsmath}
\usepackage{multirow} 
\usepackage{paralist}

\usepackage{amsthm}

\newcommand{\bra}[1] {\langle #1 |}
\newcommand{\ket}[1] {| #1 \rangle}
\newcommand{\braket}[2] {\langle #1 | #2 \rangle}

\newcommand{\Tr} {\operatorname{Tr}}

\newcommand{\moy}[1]{\langle #1 \rangle}

\newcommand{\one}{\leavevmode\hbox{\small1\normalsize\kern-.33em1}}
\newcommand{\sandwich}[3]{\mbox{$ \langle #1 | #2 | #3 \rangle $}}
\newcommand{\ba}{\begin{eqnarray}}
\newcommand{\ea}{\end{eqnarray}}

\newtheorem*{lem*}{Lemma}

\begin{document}

\title{Deriving tight error-trade-off relations \\ for approximate joint measurements of incompatible quantum observables}

\author{Cyril Branciard}
\affiliation{Centre for Engineered Quantum Systems and School of Mathematics and Physics, The University of Queensland, Brisbane,   QLD 4072, Australia}

\date{\today}

\begin{abstract}
The quantification of the ``measurement uncertainty'' aspect of Heisenberg's Uncertainty Principle---that is, the study of trade-offs between accuracy and disturbance, or between accuracies in an approximate joint measurement on two incompatible observables---has regained a lot of interest recently. Several approaches have been proposed and debated. In this paper we consider Ozawa's definitions for inaccuracies (as root-mean-square errors) in approximate joint measurements, and study how these are constrained in different cases, whether one specifies certain properties of the approximations---namely their standard deviations and/or their bias---or not. Extending our previous work [C. Branciard, Proc. Natl. Acad. Sci. U.S.A. 110, 6742 (2013)], we derive new error-trade-off relations, which we prove to be tight for pure states. We show explicitly how all previously known relations for Ozawa's inaccuracies follow from ours. While our relations are in general not tight for mixed states, we show how these can be strengthened and how tight relations can still be obtained in that case.
\end{abstract}

\maketitle

\section{Introduction}

Heisenberg's Uncertainty Principle~\cite{Heisenberg:1927ul} is undoubtedly one of the most famous characteristics of quantum theory. It is not only celebrated as a deep, fundamental feature of the quantum world by physicists, but has also entered popular culture in many (though sometimes improper!) ways~\cite{popular_culture}.

It is worth emphasizing, however, that the general term ``Uncertainty Principle'' encompasses in fact different properties of quantum theory.
It says on one hand that a quantum system cannot be prepared in such a way that a pair of incompatible---i.e. non-commuting---observables are arbitrarily well-defined (``preparation uncertainty''). Another aspect is that the measurement of one observable to a certain accuracy in general disturbs the subsequent measurement of another incompatible observable accordingly, or that one cannot in general jointly measure two incompatible observables to a perfect accuracy (``measurement uncertainty'').
Interestingly, while the latter ``measurement uncertainty'' aspect corresponds to Heisenberg's initial intuition~\cite{Heisenberg:1927ul}, the best known---and historically the first formally derived~\cite{kennard1927qmb,robertson1929tup,Schrodinger:1930aa}---uncertainty relations quantify instead the former ``preparation uncertainty'' aspect.
This has led to some confusion in their interpretation, which it is essential to clarify.

Surprisingly little work has been done in the previous century on the problem of quantifying the trade-off between (in)accuracy and disturbance, or between (in)accuracies in approximate joint measurements. The last two decades have however witnessed a renewed interest in these questions, motivated in particular by the development of quantum information science and by its implications for our understanding of quantum foundations.
An important milestone was the derivation in 2003--2004 by Ozawa~\cite{ozawa2003uvr,ozawa2004urj}, and soon after by Hall~\cite{hall2004pih}, of universally valid ``uncertainty relations'' for measurement-disturbance scenarios and for approximate joint measurements---or, as we will call them in this paper, of \emph{error-disturbance} and \emph{error-trade-off relations}. These came back in the spotlight in the last couple of years as they were experimentally tested~\cite{erhart2012edu,rozema2012vhm,Weston:2013fk,Sulyok:2013fk,Baek:2013ys,Ringbauer:2013aa,Kaneda:2013aa}, and strengthened~\cite{branciard2013ete,Weston:2013fk}.

This ``measurement uncertainty'' aspect of Heisenberg's Principle has also recently sparked an active debate in the scientific community~\cite{Busch:2013aa,rozema2013ndd,Ozawa:2013fk,Dressel:2013aa}, on the question whether Heisenberg's heuristic argument~\cite{Heisenberg:1927ul}---that the product of the inaccuracy $\epsilon_q$ of the measurement of a particle's position and the disturbance $\eta_p$ induced on its momentum is at least of the order of Planck's constant---is correct or not.
Such a relation---precisely, $\epsilon_q \, \eta_p \geq \hbar/2$---has only been formally proven a few months ago~\cite{Busch:2013aa,Dressel:2013aa}, while some authors still argue that this relation may in fact not always be satisfied~\cite{rozema2013ndd,Dressel:2013aa}. What is at stake in this controversy is the rigorous definition to be given to the inaccuracy and disturbance (in particular, whether these are state-dependent or not)---which Heisenberg was not clear about---and more generally the proper way to quantify the ``measurement uncertainty'' aspect of Heisenberg's Principle. The large number of recent publications on this subject~\cite{erhart2012edu,rozema2012vhm,Weston:2013fk,Sulyok:2013fk,Baek:2013ys,Ringbauer:2013aa,Kaneda:2013aa,branciard2013ete, Busch:2013aa,rozema2013ndd,Ozawa:2013fk,Dressel:2013aa,Fujikawa:2012aa,Fujikawa:2013ab,Di-Lorenzo:2013aa,Fujikawa:2013aa,Bastos:2013aa,Lu:2013aa,Buscemi:2013aa,Ipsen:2013aa,Busch:2013ab,Korzekwa:2013aa,Coles:2013aa} illustrates how topical these questions are, and suggests that these are still far from being settled.

Agreeing on a framework and definitions is thus a necessary starting point to derive and study error-trade-off relations. Here we consider Ozawa's (state-dependent) definitions~\cite{ozawa2003uvr,ozawa2004urj} for the inaccuracies in approximate joint measurements---for which (in the case of error-disturbance) the above relation $\epsilon_q \, \eta_p \geq \hbar/2$ can indeed be violated---and extend our previous analysis of Ref.~\cite{branciard2013ete} to characterize the allowed values of inaccuracies in a number of different cases.
The paper is structured as follows. In Section~\ref{sec_defs} we introduce the general framework for approximate joint measurements that we consider. Section~\ref{sec_err_trade_off_relations} presents the derivation of new error-trade-off relations, for cases where certain properties of the approximations---namely, their standard deviations and/or their bias with respect to the ideal measurements---are assumed to take some given values; when none of these properties are specified, we show that we obtain the relations previously derived in~\cite{branciard2013ete}. In Section~\ref{sec_altern_form} we give an alternative, equivalent form for our relations. Section~\ref{sec_tightness}, completed by the Appendix, contains the proof that our error-trade-off relations are tight for pure states (at least in certain cases for our last two relations). In Section~\ref{sec_previous_relations_follow} we show explicitly how all previously derived (non-tight) relations of Refs.~\cite{ozawa2003uvr,ozawa2004urj,hall2004pih,Weston:2013fk} follow from ours. We finish in Section~\ref{sec_mixed_states} with some considerations on the case of mixed states, and then conclude.

\section{Approximate joint measurements of incompatible observables: notations and definitions}
\label{sec_defs}

\subsection{Two incompatible observables to be measured on a given quantum state}

In this paper we consider a quantum state $\ket{\psi}$ in some finite-dimensional Hilbert space ${\cal H}$ together with two observables (Hermitian operators) $A$ and $B$ acting on ${\cal H}$, and we wish to quantify how well a joint measurement of $A$ and $B$ on $\ket{\psi}$ can be approximated.

The given observables $A$ and $B$ and the state $\ket{\psi}$ define the standard deviations\footnote{Here and throughout the paper, the notation $\moy{\cdot}$ stands for $\sandwich{\psi}{\cdot}{\psi}$ or $\sandwich{\psi,\xi}{\cdot}{\psi,\xi}$, depending on the context (or $\Tr[ \, \cdot \, \rho]$, $\Tr[ \, \cdot \, (\rho \!\otimes\! \ket{\xi}\!\bra{\xi})]$ in Section~\ref{sec_mixed_states}). Furthermore, for simplicity of notation, $A - \moy{A}$ for instance stands for $A - \moy{A} \one_{\cal H}$, where $\one_{\cal H}$ is the identity operator on ${\cal H}$.} $\Delta A = \sandwich{\psi}{(A - \moy{A})^2}{\psi}^{1/2}$ and $\Delta B = \sandwich{\psi}{(B - \moy{B})^2}{\psi}^{1/2}$, as well as the parameter
\ba
C_{\!AB} & = & \frac{1}{2i} \sandwich{\psi}{[A,B]}{\psi},
\ea
which depends on the commutator $[A,B] = AB - BA$ and quantifies the incompatibility of $A$ and $B$---as we shall see, when $C_{\!AB} \neq 0$ their perfect joint measurement on $\ket{\psi}$ is impossible.
With these notations, Roberston's well-known uncertainty relation~\cite{robertson1929tup} (which, again, quantifies the ``preparation uncertainty'' aspect of the Uncertainty Principle and not its ``measurement uncertainty'' aspect) writes
\ba
\Delta A \ \Delta B & \, \geq \, & |C_{\!AB}|. \label{eq:robertson}
\ea

We shall assume throughout the paper that\footnote{If $\Delta A \, \Delta B = 0$ our problem becomes trivial, as one of the two observables then always gives the same outcome for $\ket{\psi}$, and does not need to be actually measured; a perfect joint ``measurement'' of both observables on $\ket{\psi}$ can then be performed by simply outputting the expected outcome for that observable, and measuring the other one. Note that in such a case, $C_{\!AB} = 0$.} $\Delta A \,, \Delta B > 0$.
It will be convenient to define the reduced observables $\tilde A_0 = [A-\moy{A}]/\Delta A$, $\tilde B_0 = [B-\moy{B}]/\Delta B$, such that $\moy{\tilde A_0} = \moy{\tilde B_0} = 0$ and $\Delta \tilde A_0 = \Delta \tilde B_0 = 1$, and the reduced parameter $\tilde C_{\!AB} = \frac{C_{\!AB}}{\Delta A \, \Delta B} = {\mathrm{Im}} \, \moy{\tilde A_0 \tilde B_0}$ (where ${\mathrm{Im}}$ denotes an imaginary part).
Note that $\tilde A_0 \ket{\psi}$ and $\tilde B_0 \ket{\psi}$ are unit ket vectors, and hence $|\moy{\tilde A_0 \tilde B_0}| \leq 1$ and $\tilde C_{\!AB} = {\mathrm{Im}} \, \moy{\tilde A_0 \tilde B_0} \in [-1,1]$---as implied also by Roberston's relation above.

\subsection{Approximate joint measurements}

A general strategy for approximating the joint measurement of $A$ and $B$ is to actually measure two compatible---i.e. \emph{commuting}---observables ${\cal A}$ and ${\cal B}$, which are taken to approximate $A$ and $B$, respectively. In full generality these can be measured on an extended Hilbert space, i.e. on the state $\ket{\psi,\xi} = \ket{\psi} \otimes \ket{\xi} \in {\cal H} \otimes {\cal K}$, where $\ket{\xi}$ is the state of an ancillary system in some ancillary Hilbert space\footnote{Another way to present the general case is to consider a Positive Operator Valued Measure (POVM) instead of projective measurements ${\cal A}$ and ${\cal B}$; see the Supporting Information of Ref.~\cite{branciard2013ete} for a detailed treatment from this perspective.} ${\cal K}$. Following Ozawa~\cite{Ozawa:1991aa,ozawa2003uvr,ozawa2004urj}, we quantify here the \emph{inaccuracies} in the approximations of $A$ and $B$ by the \emph{root-mean-square errors}
\ba
\epsilon_{\cal A} &=& \sandwich{\psi,\xi}{\,({\cal A} - A)^2\,}{\psi,\xi}^{1/2} , \label{eq:def_epsA} \\
\epsilon_{\cal B} &=& \sandwich{\psi,\xi}{\,({\cal B} - B)^2\,}{\psi,\xi}^{1/2} , \label{eq:def_epsB}
\ea
where for simplicity $A$ and $B$ are used as shorthand notations for $A \otimes \one_{\cal K}$ and $B \otimes \one_{\cal K}$, respectively.

The approximations ${\cal A}$ and ${\cal B}$ also define (together with $\ket{\xi}$) the standard deviations
$\Delta {\cal A} = \sandwich{\psi,\xi}{({\cal A} - \moy{{\cal A}})^2}{\psi,\xi}^{1/2}$ and $\Delta {\cal B} = \sandwich{\psi,\xi}{({\cal B} - \moy{{\cal B}})^2}{\psi,\xi}^{1/2}$, and the (first moment) biases $\delta_{\cal A} = \moy{{\cal A} - A}$ and $\delta_{\cal B} = \moy{{\cal B} - B}$.
These quantities are related through
\ba
\epsilon_{\cal A}^2 - \delta_{\cal A}^2 &=& \Delta A^2 + \Delta {\cal A}^2 - 2 \, {\mathrm{Re}} \, \moy{(A{-}\moy{A})({\cal A}{-}\moy{{\cal A}})} , \quad \nonumber \\
 &=& \Delta A^2 + \Delta {\cal A}^2 - 2 \, \Delta A \ {\mathrm{Re}} \, \moy{\tilde A_0{\cal A}} , \quad \label{eq:relation_eps_delta_Deltas_A} \\[2mm]
\epsilon_{\cal B}^2 - \delta_{\cal B}^2 &=& \Delta B^2 + \Delta {\cal B}^2 - 2 \, {\mathrm{Re}} \, \moy{(B{-}\moy{B})({\cal B}{-}\moy{{\cal B}})} , \nonumber \\
 &=& \Delta B^2 + \Delta {\cal B}^2 - 2 \, \Delta B \ {\mathrm{Re}} \, \moy{\tilde B_0{\cal B}} , \quad \label{eq:relation_eps_delta_Deltas_B}
\ea
where ${\mathrm{Re}}$ indicates a real part.
Using the fact that (from the Cauchy-Schwartz inequality) $\big| \moy{(A{-}\moy{A})({\cal A}{-}\moy{{\cal A}})} \big| \leq \Delta A \, \Delta {\cal A}$, and similarly for $B$ and ${\cal B}$, it follows in particular that\footnote{Noting that $\epsilon_{\cal A}^2{-}\delta_{\cal A}^2 = \Delta [{\cal A}{-}A]^2$ and $\epsilon_{\cal B}^2{-}\delta_{\cal B}^2 = \Delta [{\cal B}{-}B]^2$, Eqs.~(\ref{eq:bnds_eps_delta_Deltas_A}--\ref{eq:bnds_eps_delta_Deltas_B}) are in fact nothing but triangle inequalities for variances.}
\ba
(\Delta A - \Delta{\cal A})^2 \ \leq & \ \ \epsilon_{\cal A}^2 - \delta_{\cal A}^2 \ \ & \leq \ (\Delta A + \Delta{\cal A})^2 , \label{eq:bnds_eps_delta_Deltas_A} \\
(\Delta B - \Delta{\cal B})^2 \ \leq & \ \ \epsilon_{\cal B}^2 - \delta_{\cal B}^2 \ \ & \leq \ (\Delta B + \Delta{\cal B})^2 . \label{eq:bnds_eps_delta_Deltas_B}
\ea

\medskip

Note that while the above framework is presented for the scenario of approximate joint measurements, the case of measurement-disturbance---where an approximate measurement of $A$ disturbs a subsequent measurement of $B$---can also be cast into the same framework~\cite{Ozawa:1991aa,ozawa2003uvr}. In that case the \emph{inaccuracy} (root-mean-square error) $\epsilon_{\cal B}$ is interpreted as the \emph{disturbance} $\eta_{\cal B}$ on $B$, with the same definition~\eqref{eq:def_epsB}. All the error-trade-off relations we shall derive, which bound the possible values of $(\epsilon_{\cal A},\epsilon_{\cal B})$, hence also hold as a particular case for $(\epsilon_{\cal A},\eta_{\cal B})$ and can thus also be interpreted as error-disturbance relations. An important difference between the two scenarios, however, is that in the measurement-disturbance case ${\cal B}$ is forced to have the same spectrum as $B$; this may in general impose stronger constraints, leading to stronger error-disturbance relations~\cite{branciard2013ete} (see Subsection~\ref{subsec_relations_same_spectrum}).

\subsection{Fixed versus variable parameters}
\label{subsec_fixed_vs_variable}

As defined above, the parameters $\Delta A, \Delta B$, $\moy{A}, \moy{B}$ and $\tilde C_{\!AB}$ depend on the given state $\ket{\psi}$ and the given observables $A$ and $B$ under consideration. These will be considered fixed throughout the paper.

On the other hand, the parameters $\Delta {\cal A}, \Delta {\cal B}$, $\delta_{\cal A}, \delta_{\cal B}$ (or equivalently $\moy{{\cal A}}, \moy{{\cal B}}$), $\epsilon_{\cal A}$ and $\epsilon_{\cal B}$ depend on the particular choice of approximation strategy---namely, on ${\cal A}$ and ${\cal B}$. These will be considered as variable parameters. Depending on the situation under study, we shall consider cases where their values are specified---e.g. the values of $\Delta {\cal A}$ and $\Delta {\cal B}$ are specified in Subsection~\ref{subsec_relation_g} below, and the error-trade-off relation derived there holds when the approximations indeed give the specified values; similarly, $\delta_{\cal A}$ and $\delta_{\cal B}$ are assumed to be given in Subsection~\ref{subsec_relation_h}---or we shall let them free---in which case these can be optimized to minimize for instance the values of $(\epsilon_{\cal A},\epsilon_{\cal B})$.

\section{Error-trade-off relations}
\label{sec_err_trade_off_relations}

We derive in this section different error-trade-off relations that restrict the allowed values of $(\epsilon_{\cal A},\epsilon_{\cal B})$, as a function of different parameters ---namely, whether or not one may specify the values of $\Delta {\cal A}, \Delta {\cal B}$ and/or $\delta_{\cal A}, \delta_{\cal B}$---and whether or not an additional assumption---the ``same-spectrum assumption'', see subsection~\ref{subsec_relations_same_spectrum} below---is imposed.
All these relations will be shown to follow from our first relation~\eqref{eq:relation_f} below.

\subsection{An error-trade-off relation \\ for specified values of $\Delta {\cal A}, \delta_{\cal A}$, $\Delta {\cal B}$ and $\delta_{\cal B}$}
\label{subsec_relation_f}

The first relation we derive holds for the case where the standard deviations $\Delta {\cal A}, \Delta {\cal B}$ and the biases $\delta_{\cal A}, \delta_{\cal B}$ are specified.
We assume that $\Delta {\cal A}, \Delta {\cal B} \neq 0$, and define
\ba
f_{\Delta {\cal A}, \delta_{\cal A}}(\epsilon_{\cal A}) & = & \sqrt{1{-}\Big( \frac{\Delta A^2 + \Delta {\cal A}^2 - (\epsilon_{\cal A}^2 - \delta_{\cal A}^2)
}{2 \, \Delta A \, \Delta {\cal A}} \Big)^{\!2} }, \label{def_fA} \\
f_{\Delta {\cal B}, \delta_{\cal B}}(\epsilon_{\cal B}) & = & \sqrt{1{-}\Big( \frac{\Delta B^2 + \Delta {\cal B}^2  - (\epsilon_{\cal B}^2 - \delta_{\cal B}^2)}{2 \, \Delta B \, \Delta {\cal B}} \Big)^{\!2} }. \qquad \label{def_fB}
\ea
Note that Eqs.~(\ref{eq:bnds_eps_delta_Deltas_A}--\ref{eq:bnds_eps_delta_Deltas_B}) ensure for instance that $\big| \frac{\Delta A^2 + \Delta {\cal A}^2 - (\epsilon_{\cal A}^2 - \delta_{\cal A}^2)
}{2 \, \Delta A \, \Delta {\cal A}} \big| \leq 1$, and that the above square roots are real.

Our first error-trade-off relation states that for any given values of $\Delta {\cal A} > 0, \delta_{\cal A}, \Delta {\cal B} > 0$ and $\delta_{\cal B}$ (and whatever the values of $\Delta A, \Delta B$ and $\tilde C_{\!AB}$, which---we recall---are fixed parameters in our study once $A,B$ and $\ket{\psi}$ are fixed), the inaccuracies $(\epsilon_{\cal A}, \epsilon_{\cal B})$ are bound to satisfy
\ba
& \hspace{-4cm} f_{\Delta {\cal A}, \delta_{\cal A}}(\epsilon_{\cal A})^2 + f_{\Delta {\cal B}, \delta_{\cal B}}(\epsilon_{\cal B})^2 \nonumber \\
&+ 2 \sqrt{1 - \tilde C_{\!AB}^2} \ f_{\Delta {\cal A}, \delta_{\cal A}}(\epsilon_{\cal A}) \, f_{\Delta {\cal B}, \delta_{\cal B}}(\epsilon_{\cal B}) \ \geq \ \tilde C_{\!AB}^2 . \quad \label{eq:relation_f}
\ea

The proof follows similar lines as the proofs of Ref.~\cite{branciard2013ete}.
It makes use of the following geometric Lemma, introduced and proven previously in~\cite{branciard2013ete}:
\begin{lem*}
Let $\hat a, \hat b$ be two unit vectors of a Euclidean space ${\cal E}$, and let us denote by $\chi = \hat a \cdot \hat b$ their scalar product. For any two orthogonal unit vectors $\hat x$ and $\hat y$ of ${\cal E}$, defining
$a_\perp = \sqrt{1-(\hat a \cdot \hat x)^2}$ and $b_\perp = \sqrt{1-(\hat b \cdot \hat y)^2}$, one has
\ba
a_\perp^2 + b_\perp^2 + 2 \sqrt{1 - \chi^2} \ a_\perp \, b_\perp \ \geq \ \chi^2. \label{eq:geom_lemma_ineq}
\ea
\end{lem*}
\noindent(Note furthermore that a necessary condition for inequality~\eqref{eq:geom_lemma_ineq} to be saturated is that $\hat a, \hat b, \hat x$ and $\hat y$ are coplanar~\cite{branciard2013ete}.)

\begin{proof}[Proof of Eq.~\eqref{eq:relation_f}]

Assuming $\Delta {\cal A}, \Delta {\cal B} > 0$, let us define $\tilde {\cal A}_0 = [{\cal A}-\moy{{\cal A}}]/\Delta {\cal A}$, $\tilde {\cal B}_0 = [{\cal B}-\moy{{\cal B}}]/\Delta {\cal B}$, and the (normalized) ket vectors
\ba
\ket{a} = \tilde A_0 \ket{\psi,\xi}, & \quad & \ket{b} = \tilde B_0 \ket{\psi,\xi}, \quad \nonumber \\[1mm]
\ket{x} = \tilde {\cal A}_0 \ket{\psi,\xi}, & \quad & \ket{y} = \tilde {\cal B}_0 \ket{\psi,\xi} \nonumber
\ea
(where, as previously, $\tilde A_0$ and $\tilde B_0$ are used as shorthand notations for $\tilde A_0 \otimes \one_{\cal K}$ and $\tilde B_0 \otimes \one_{\cal K}$).
By writing these vectors in any orthonormal basis of ${\cal H} \otimes {\cal K}$ (e.g., the common eigenbasis of ${\cal A}$ and ${\cal B}$),
one can define the following real vectors:
\ba
\hat a = \left(
\begin{array}{c}
{\mathrm{Re}}\ket{a} \\
{\mathrm{Im}}\ket{a}
\end{array}
\right), \quad
\hat b = \left(
\begin{array}{c}
{\mathrm{Im}}\ket{b} \\
{-\mathrm{Re}}\ket{b}
\end{array}
\right), \nonumber \\[1mm]
\hat x = \left(
\begin{array}{c}
{\mathrm{Re}}\ket{x} \\
{\mathrm{Im}}\ket{x}
\end{array}
\right), \quad
\hat y = \left(
\begin{array}{c}
{\mathrm{Im}}\ket{y} \\
{-\mathrm{Re}}\ket{y}
\end{array}
\right). \nonumber
\ea

One then has
\ba
\|\hat a\|^2 & = & ({\mathrm{Re}}\ket{a})^{\!\top} \!\!\cdot\! ({\mathrm{Re}}\ket{a}) + ({\mathrm{Im}}\ket{a})^{\!\top} \!\!\cdot\! ({\mathrm{Im}}\ket{a}) = \braket{a}{a} = 1, \nonumber \\[-1mm] \nonumber \\
\|\hat b\|^2 & = & \|\hat x\|^2 \,=\, \|\hat y\|^2 \,=\, 1, \nonumber \\[2mm]
\hat a \cdot \hat b &=& ({\mathrm{Re}}\ket{a})^{\!\top} \!\!\cdot ({\mathrm{Im}}\ket{b}) - ({\mathrm{Im}}\ket{a})^{\!\top} \!\!\cdot ({\mathrm{Re}}\ket{b}) = {\mathrm{Im}}\,\braket{a}{b} \nonumber \\
 &=& {\mathrm{Im}}\,\moy{\tilde A_0\tilde B_0} \, = \, \tilde C_{\!AB}, \qquad \nonumber \\[2mm]
\hat x \cdot \hat y &=& {\mathrm{Im}}\,\moy{\tilde {\cal A}_0\tilde {\cal B}_0} = \frac{1}{2i} \frac{\sandwich{\psi,\xi}{[{\cal A}, {\cal B}]}{\psi,\xi}}{\Delta {\cal A} \, \Delta {\cal B}} = 0. \nonumber
\ea
Hence, the real vectors $\hat a, \hat b, \hat x, \hat y$ satisfy the assumptions of the geometric Lemma above, with $\chi = \tilde C_{\!AB}$. Furthermore,
\ba
\hat a \cdot \hat x &=& ({\mathrm{Re}}\ket{a})^{\!\top} \!\!\cdot ({\mathrm{Re}}\ket{x}) + ({\mathrm{Im}}\ket{a})^{\!\top} \!\!\cdot ({\mathrm{Im}}\ket{x}) = {\mathrm{Re}}\,\braket{a}{x} \nonumber \\
& = & \frac{{\mathrm{Re}}\,\moy{(A{-}\moy{A})({\cal A}{-}\moy{{\cal A}})}}{\Delta A \, \Delta {\cal A}} = \frac{\Delta A^2{+}\Delta {\cal A}^2{-}(\epsilon_{\cal A}^2{-}\delta_{\cal A}^2)}{2 \, \Delta A \, \Delta {\cal A}}, \nonumber
\ea
where the last equality follows from Eq.~\eqref{eq:relation_eps_delta_Deltas_A}, and thus
\ba
a_\perp &=& \sqrt{1-(\hat a \cdot \hat x)^2} = f_{\Delta {\cal A}, \delta_{\cal A}}(\epsilon_{\cal A}). \nonumber
\ea

Similarly, one finds that $b_\perp = f_{\Delta {\cal B}, \delta_{\cal B}}(\epsilon_{\cal B})$.
Eq.~\eqref{eq:relation_f} then directly follows from inequality~\eqref{eq:geom_lemma_ineq} of the Lemma.

\end{proof}

Relation~\eqref{eq:relation_f} lower-bounds the possible values of $(f_{\Delta {\cal A}, \delta_{\cal A}}(\epsilon_{\cal A}), f_{\Delta {\cal B}, \delta_{\cal B}}(\epsilon_{\cal B}))$ for some specified values of $\Delta {\cal A}, \delta_{\cal A}$, $\Delta {\cal B}$ and $\delta_{\cal B}$.
Note that $f_{\Delta {\cal A}, \delta_{\cal A}}(\epsilon_{\cal A})$ increases with $\epsilon_{\cal A}$ for $\epsilon_{\cal A}^2 \leq \Delta A^2 + \Delta {\cal A}^2 + \delta_{\cal A}^2$ and decreases for $\epsilon_{\cal A}^2 \geq \Delta A^2 + \Delta {\cal A}^2 + \delta_{\cal A}^2$ (and similarly for $f_{\Delta {\cal B}, \delta_{\cal B}}(\epsilon_{\cal B})$). Hence, relation~\eqref{eq:relation_f} bounds $\epsilon_{\cal A}$ and $\epsilon_{\cal B}$ from both below and above.
Note furthermore that due to~(\ref{eq:bnds_eps_delta_Deltas_A}--\ref{eq:bnds_eps_delta_Deltas_B}), there are absolute lower- and upper-bounds on $\epsilon_{\cal A}$ and $\epsilon_{\cal B}$; for instance, if $\Delta {\cal A}$ and $\delta_{\cal A}$ are such that $(\Delta A - \Delta{\cal A})^2 + \delta_{\cal A}^2 > 0$, then whatever $\epsilon_{\cal B}$, $\epsilon_{\cal A}$ cannot be 0.

\medskip

In the case where $\Delta {\cal A} = 0$, one finds (e.g. from~\eqref{eq:bnds_eps_delta_Deltas_A}) that $\epsilon_{\cal A}^2 = \Delta A^2 + \delta_{\cal A}^2$ is fixed, and $\epsilon_{\cal B}$ is then only\footnote{Note that when $\Delta {\cal A} = 0$, ${\cal A}$ can be taken to be proportional to the identity (it then does not actually need to be measured), which commutes with any observable ${\cal B}$; there is thus no additional constraint on ${\cal B}$.} bounded by Eq.~\eqref{eq:bnds_eps_delta_Deltas_B}. The constraint on $(\epsilon_{\cal A}, \epsilon_{\cal B})$ thus writes
\ba
\left\{
\begin{array}{l}
\epsilon_{\cal A}^2 = \Delta A^2 + \delta_{\cal A}^2 \\[1mm]
(\Delta B - \Delta{\cal B})^2 + \delta_{\cal B}^2 \ \leq \ \epsilon_{\cal B}^2 \ \leq \ (\Delta B + \Delta{\cal B})^2 + \delta_{\cal B}^2 . \qquad
\end{array}
\right.
\label{eq:relation_f_limit_DA_0}
\ea
Similarly, in the case where $\Delta {\cal B} = 0$,
\ba
\left\{
\begin{array}{l}
\epsilon_{\cal B}^2 = \Delta B^2 + \delta_{\cal B}^2 \\[1mm]
(\Delta A - \Delta{\cal A})^2 + \delta_{\cal A}^2 \ \leq \ \epsilon_{\cal A}^2 \ \leq \ (\Delta A + \Delta{\cal A})^2 + \delta_{\cal A}^2 . \qquad
\end{array}
\right.
\label{eq:relation_f_limit_DB_0}
\ea
These constraints indeed correspond to the limits of the constraint~\eqref{eq:relation_f} when $\Delta {\cal A} \to 0$ and $\Delta {\cal B} \to 0$, respectively.

\subsection{An error-trade-off relation \\ for specified values of $\Delta {\cal A}$ and $\Delta {\cal B}$}
\label{subsec_relation_g}

None of the previously derived error-trade-off relations~\cite{ozawa2003uvr,ozawa2004urj,hall2004pih,Weston:2013fk,branciard2013ete} involved the biases $\delta_{\cal A}$ or $\delta_{\cal B}$ explicitly. 
When these parameters are not specified, one can optimize them and derive, from~\eqref{eq:relation_f}, error-trade-off relations which do not include them, as follows.

Note that when $\delta_{\cal A}$ and $\delta_{\cal A}$ are left free, the only constraints on $\epsilon_{\cal A}$ and $\epsilon_{\cal B}$ that follow from~(\ref{eq:bnds_eps_delta_Deltas_A}--\ref{eq:bnds_eps_delta_Deltas_B}) are $\epsilon_{\cal A} \geq |\Delta A - \Delta{\cal A}|$ and $\epsilon_{\cal B} \geq |\Delta B - \Delta{\cal B}|$ (which ensure in particular that $\frac{\Delta A^2 + \Delta {\cal A}^2 - \epsilon_{\cal A}^2}{2 \, \Delta A \, \Delta {\cal A}} \leq 1$ and $\frac{\Delta B^2 + \Delta {\cal B}^2 - \epsilon_{\cal B}^2}{2 \, \Delta B \, \Delta {\cal B}} \leq 1$): $\epsilon_{\cal A}$ and $\epsilon_{\cal B}$ still have an absolute lower-bound, but are not upper-bounded any more.

Using the fact that $\big| \Delta A^2 + \Delta {\cal A}^2 - (\epsilon_{\cal A}^2 - \delta_{\cal A}^2) \big| \geq \max [\Delta A^2 + \Delta {\cal A}^2 - \epsilon_{\cal A}^2,0] \geq 0$, it follows that for all $\Delta {\cal A} > 0, \delta_{\cal A}$ and $\epsilon_{\cal A}$ satisfying~\eqref{eq:bnds_eps_delta_Deltas_A},
\ba
&& f_{\Delta {\cal A}, \delta_{\cal A}}(\epsilon_{\cal A}) \ \leq \ g_{\Delta {\cal A}}(\epsilon_{\cal A}) , \quad \text{with} \label{fA_leq_gA} \\[1mm]
&& g_{\Delta {\cal A}}(\epsilon_{\cal A}) = \sqrt{1 - \max \Big[ \frac{\Delta A^2 + \Delta {\cal A}^2 - \epsilon_{\cal A}^2}{2 \, \Delta A \, \Delta {\cal A}}, 0 \Big]^2 } , \qquad \label{def_gA}
\ea
and with equality in~\eqref{fA_leq_gA} for $\delta_{\cal A}^2 = \max[0,\epsilon_{\cal A}^2 - (\Delta A^2 + \Delta {\cal A}^2)]$.
Similarly, one can show that for all $\Delta {\cal B} > 0, \delta_{\cal B}$ and $\epsilon_{\cal B}$ satisfying~\eqref{eq:bnds_eps_delta_Deltas_B},
\ba
&& f_{\Delta {\cal B}, \delta_{\cal B}}(\epsilon_{\cal B}) \ \leq \ g_{\Delta {\cal B}}(\epsilon_{\cal B}) , \quad \text{with} \label{fB_leq_gB} \\[1mm]
&& g_{\Delta {\cal B}}(\epsilon_{\cal B}) = \sqrt{1 - \max \Big[ \frac{\Delta B^2 + \Delta {\cal B}^2 - \epsilon_{\cal B}^2}{2 \, \Delta B \, \Delta {\cal B}}, 0 \Big]^2 } , \qquad  \label{def_gB}
\ea
with equality in~\eqref{fB_leq_gB} for $\delta_{\cal B}^2 = \max[0,\epsilon_{\cal B}^2 - (\Delta B^2 + \Delta {\cal B}^2)]$.

Using Eqs.~\eqref{fA_leq_gA} and~\eqref{fB_leq_gB}, it follows from our previous relation~\eqref{eq:relation_f} that for any given values of $\Delta {\cal A} > 0$ and $\Delta {\cal B} > 0$,
\ba
& \hspace{-4cm} g_{\Delta {\cal A}}(\epsilon_{\cal A})^2 + g_{\Delta {\cal B}}(\epsilon_{\cal B})^2 \nonumber \\
&+ 2 \sqrt{1 - \tilde C_{\!AB}^2} \ g_{\Delta {\cal A}}(\epsilon_{\cal A}) \, g_{\Delta {\cal B}}(\epsilon_{\cal B}) \ \geq \ \tilde C_{\!AB}^2 . \quad \label{eq:relation_g}
\ea
This relation lower-bounds the possible values of $(g_{\Delta {\cal A}}(\epsilon_{\cal A}), g_{\Delta {\cal B}}(\epsilon_{\cal B}))$, for some specified values of $\Delta {\cal A}$ and $\Delta {\cal B}$. Now, $g_{\Delta {\cal A}}(\epsilon_{\cal A})$ and $g_{\Delta {\cal B}}(\epsilon_{\cal B})$ only increase with $\epsilon_{\cal A}$ and $\epsilon_{\cal B}$. Hence, the above relation also only bounds the allowed values of $(\epsilon_{\cal A}, \epsilon_{\cal B})$ from below (with absolute lower-bounds of $|\Delta A - \Delta{\cal A}|$ and $|\Delta B - \Delta{\cal B}|$, resp.); as emphasized above, these are no longer bounded from above. This relation can be compared to those of Refs~\cite{hall2004pih,Weston:2013fk}, which also involve the parameters $\Delta {\cal A}$ and $\Delta {\cal B}$ (see Subsections~\ref{subsec_hall} and~\ref{subsec_weston} below).

\medskip

In the case where $\Delta {\cal A} = 0$ or $\Delta {\cal B} = 0$, the constraint on $(\epsilon_{\cal A}, \epsilon_{\cal B})$ is simply
\ba
\left\{
\begin{array}{l}
\epsilon_{\cal A}^2 \geq \Delta A^2 \\[1mm]
\epsilon_{\cal B}^2 \geq (\Delta B - \Delta{\cal B})^2
\end{array}
\right. \ \text{or} \
\left\{
\begin{array}{l}
\epsilon_{\cal A}^2 \geq (\Delta A - \Delta{\cal A})^2 \\[1mm]
\epsilon_{\cal B}^2 \geq \Delta B^2
\end{array}
\right. , \quad
\label{eq:relation_g_limit_DA_0_DB_0}
\ea
which indeed correspond to the limits of the constraint~\eqref{eq:relation_g} when $\Delta {\cal A} \to 0$ or $\Delta {\cal B} \to 0$, respectively.

\subsection{An error-trade-off relation \\ for specified values of $\delta_{\cal A}$ and $\delta_{\cal B}$}
\label{subsec_relation_h}

Instead of relaxing and letting the values of $\delta_{\cal A}, \delta_{\cal B}$ free as above, one may relax the values of $\Delta {\cal A}, \Delta {\cal B}$ and optimize over these to derive, from~\eqref{eq:relation_f}, error-trade-off relations which do not include the latter.
Note that when $\Delta {\cal A}$ and $\Delta {\cal B}$ are left free, the only constraints on $\epsilon_{\cal A}$ and $\epsilon_{\cal B}$ that follow from~(\ref{eq:bnds_eps_delta_Deltas_A}--\ref{eq:bnds_eps_delta_Deltas_B}) are $\epsilon_{\cal A} \geq \delta_{\cal A}$ and $\epsilon_{\cal B} \geq \delta_{\cal B}$; as before, $\epsilon_{\cal A}$ and $\epsilon_{\cal B}$ still have an absolute lower-bound, but are not upper-bounded any more.

By looking for the maximum of $f_{\Delta {\cal A}, \delta_{\cal A}}(\epsilon_{\cal A})$ as $\Delta {\cal A}$ varies\footnote{Formally, one could also look for the maximum of $f_{\Delta {\cal A}, \delta_{\cal A}}(\epsilon_{\cal A})$ as $\Delta A$ varies and of $f_{\Delta {\cal B}, \delta_{\cal B}}(\epsilon_{\cal B})$ as $\Delta B$ varies, and derive a similar relation to~\eqref{eq:relation_h}, with $\Delta A$ and $\Delta B$ replaced by $\Delta {\cal A}$ and $\Delta {\cal B}$ in the definitions of $h_{\delta_{\cal A}}(\epsilon_{\cal A})$ and $h_{\delta_{\cal B}}(\epsilon_{\cal B})$ (but not in the definition of $\tilde C_{\!AB}$!)---and a similar relation to~\eqref{eq:relation_PNAS} as well, with $\tilde \epsilon_{\cal A}$ and $\tilde \epsilon_{\cal B}$ replaced by $\epsilon_{\cal A} / \Delta {\cal A}$ and $\epsilon_{\cal B} / \Delta {\cal B}$, resp. However, as explained in Subsection~\ref{subsec_fixed_vs_variable}, this is not in the spirit of our approach to vary $\Delta A$ and $\Delta B$, and the error-trade-off relation thus obtained would in general not be tight.} between $\big| \Delta A - \sqrt{\epsilon_{\cal A}^2 - \delta_{\cal A}^2} \big|$ and $\Delta A + \sqrt{\epsilon_{\cal A}^2 - \delta_{\cal A}^2}$ (so as to satisfy~\eqref{eq:bnds_eps_delta_Deltas_A}), while $\delta_{\cal A}$ and $\epsilon_{\cal A}$ are kept fixed, one can show that for all $\Delta {\cal A} > 0, \delta_{\cal A}$ and $\epsilon_{\cal A}$ satisfying~\eqref{eq:bnds_eps_delta_Deltas_A},
\ba
&& f_{\Delta {\cal A}, \delta_{\cal A}}(\epsilon_{\cal A}) \ \leq \ h_{\delta_{\cal A}}(\epsilon_{\cal A}) , \quad \text{with} \label{fA_leq_hA} \\[1mm]
&& h_{\delta_{\cal A}}(\epsilon_{\cal A}) = \frac{\sqrt{\epsilon_{\cal A}^2 - \delta_{\cal A}^2}}{\Delta A}, \qquad \label{def_hA}
\ea
with equality in~\eqref{fA_leq_hA} if $\epsilon_{\cal A}^2 - \delta_{\cal A}^2 \leq \Delta A^2$ and $\Delta {\cal A} = \sqrt{\Delta A^2 - (\epsilon_{\cal A}^2 - \delta_{\cal A}^2)}$. Note that contrary to $f_{\Delta {\cal A}, \delta_{\cal A}}(\epsilon_{\cal A})$ and $g_{\Delta {\cal A}}(\epsilon_{\cal A})$, $h_{\delta_{\cal A}}(\epsilon_{\cal A})$ is well defined when $\Delta {\cal A} = 0$, in which case it takes the value 1 (as from~\eqref{eq:bnds_eps_delta_Deltas_A}, $\Delta {\cal A} = 0$ implies $\epsilon_{\cal A}^2 - \delta_{\cal A}^2 = \Delta A^2$).

Similarly, one can show that for all $\Delta {\cal B} > 0, \delta_{\cal B}$ and $\epsilon_{\cal B}$ satisfying~\eqref{eq:bnds_eps_delta_Deltas_B},
\ba
&& f_{\Delta {\cal B}, \delta_{\cal B}}(\epsilon_{\cal B}) \ \leq \ h_{\delta_{\cal B}}(\epsilon_{\cal B}) , \quad \text{with} \label{fB_leq_hB} \\[1mm]
&& h_{\delta_{\cal B}}(\epsilon_{\cal B}) = \frac{\sqrt{\epsilon_{\cal B}^2 - \delta_{\cal B}^2}}{\Delta B} , \qquad \label{def_hB}
\ea
with equality in~\eqref{fB_leq_hB} if $\epsilon_{\cal B}^2 - \delta_{\cal B}^2 \leq \Delta B^2$ and $\Delta {\cal B} = \sqrt{\Delta B^2 - (\epsilon_{\cal B}^2 - \delta_{\cal B}^2)}$. Again, $h_{\delta_{\cal B}}(\epsilon_{\cal B})$ is well defined when $\Delta {\cal B} = 0$, in which case it cases the value 1.

Using Eqs.~\eqref{fA_leq_hA} and~\eqref{fB_leq_hB}, it follows from our previous relation~\eqref{eq:relation_f} that for any given values of $\delta_{\cal A}$ and $\delta_{\cal A}$,
\ba
& \hspace{-4cm} h_{\delta_{\cal A}}(\epsilon_{\cal A})^2 + h_{\delta_{\cal B}}(\epsilon_{\cal B})^2 \nonumber \\
&+ 2 \sqrt{1 - \tilde C_{\!AB}^2} \ h_{\delta_{\cal A}}(\epsilon_{\cal A}) \, h_{\delta_{\cal B}}(\epsilon_{\cal B}) \ \geq \ \tilde C_{\!AB}^2 . \quad \label{eq:relation_h}
\ea
(Strictly speaking, this relation has so far been proven in the case where $\Delta {\cal A}, \Delta {\cal B} > 0$; given the above remarks, when $\Delta {\cal A} = 0$ or $\Delta {\cal B} = 0$ the relation also trivially holds.)

Relation~\eqref{eq:relation_h} lower-bounds the possible values of $(h_{\delta_{\cal A}}(\epsilon_{\cal A}), h_{\delta_{\cal B}}(\epsilon_{\cal B}))$, for some specified values of $\delta_{\cal A}$ and $\delta_{\cal B}$. Now, as it was the case with $g_{\Delta {\cal A}}(\epsilon_{\cal A})$ and $g_{\Delta {\cal B}}(\epsilon_{\cal B})$, $h_{\delta_{\cal A}}(\epsilon_{\cal A})$ and $h_{\delta_{\cal B}}(\epsilon_{\cal B})$ only increase with $\epsilon_{\cal A}$ and $\epsilon_{\cal B}$. Hence, the above relation also only bounds the allowed values of $(\epsilon_{\cal A}, \epsilon_{\cal B})$ from below (with absolute lower-bounds $\delta_{\cal A}$ and $\delta_{\cal B}$, resp.); as emphasized above, these are again not bounded from above.

\subsection{A general error-trade-off relation \\ for any values of $\Delta {\cal A}, \delta_{\cal A}$, $\Delta {\cal B}$ or $\delta_{\cal B}$}
\label{subsec_relation_eps}

From~\eqref{eq:relation_h}, one can now easily derive an error-trade-off relation that involves neither $\Delta {\cal A}$, $\Delta {\cal B}$, nor $\delta_{\cal A}$, $\delta_{\cal B}$, as e.g. that of Ozawa~\cite{ozawa2003uvr}.

It is indeed trivial to note that for all $\delta_{\cal A}$, $\epsilon_{\cal A} \ (\geq \delta_{\cal A})$, $\delta_{\cal B}$ and $\epsilon_{\cal B} \ (\geq \delta_{\cal B})$,
\ba
h_{\delta_{\cal A}}(\epsilon_{\cal A}) \leq \frac{\epsilon_{\cal A}}{\Delta A} \quad \text{and} \quad
h_{\delta_{\cal B}}(\epsilon_{\cal B}) \leq \frac{\epsilon_{\cal B}}{\Delta B},
\ea
with equality for $\delta_{\cal A} = 0$ and $\delta_{\cal B} = 0$, resp.
Defining
\ba
\tilde \epsilon_{\cal A} = \frac{\epsilon_{\cal A}}{\Delta A} \quad \text{and} \quad \tilde \epsilon_{\cal B} = \frac{\epsilon_{\cal B}}{\Delta B},
\ea
it then follows from~\eqref{eq:relation_h}\footnote{Note that~\eqref{eq:relation_PNAS} can also be proven to follow from~\eqref{eq:relation_g}, after showing that (for $\Delta {\cal A}, \Delta {\cal B} > 0$) $g_{\Delta {\cal A}}(\epsilon_{\cal A}) \leq \tilde \epsilon_{\cal A}$ and $g_{\Delta {\cal B}}(\epsilon_{\cal B}) \leq \tilde \epsilon_{\cal B}$.} that
\ba
\tilde \epsilon_{\cal A}^2 + \tilde \epsilon_{\cal B}^2 + 2 \sqrt{1 - \tilde C_{\!AB}^2} \ \tilde \epsilon_{\cal A} \ \tilde \epsilon_{\cal B} \ \geq \ \tilde C_{\!AB}^2 ,\label{eq:relation_PNAS}
\ea
which is precisely the error-trade-off relation recently introduced in Ref.~\cite{branciard2013ete}\footnote{The proof of relation~\eqref{eq:relation_PNAS} given in~\cite{branciard2013ete} is based on a corollary of the geometric Lemma of Subsection~\ref{subsec_relation_f} and, although similar, is somewhat more direct than the present derivation. However, our approach here allows us to explicitly show how all our relations follow from one another.}, and experimentally tested in~\cite{Ringbauer:2013aa,Kaneda:2013aa}.
This relation lower-bounds the possible values of $(\epsilon_{\cal A}, \epsilon_{\cal B})$ as a function of the fixed parameters $\Delta A, \Delta B$, and $\tilde C_{\!AB}$ of the problem, whatever the values of the (variable) parameters $\Delta {\cal A}, \delta_{\cal A}$, $\Delta {\cal B}$ and $\delta_{\cal B}$.

Note that contrary to the previous cases, when $\Delta {\cal A}, \delta_{\cal A}$, $\Delta {\cal B}$ and $\delta_{\cal B}$ are free to take any values there are no absolute lower bounds on $\epsilon_{\cal A}$ and $\epsilon_{\cal B}$, which can (individually) take any non-negative value (there are no upper-bounds neither); for instance, $\epsilon_{\cal A}$ can always be zero if $\epsilon_{\cal B}$ is large enough.

\subsection{With binary, $\pm 1$-valued observables \\ and the same-spectrum assumption}
\label{subsec_relations_same_spectrum}

Our previous relations~\eqref{eq:relation_f}, \eqref{eq:relation_g}, \eqref{eq:relation_h}, \eqref{eq:relation_PNAS} are valid (and tight, as we will show in Section~\ref{sec_tightness} and in the Appendix) for any state $\ket{\psi}$ of any (finite-dimensional) Hilbert space, any pair of observables $A$ and $B$, and any commuting approximate observables ${\cal A}$ and ${\cal B}$---in particular, we did not assume that ${\cal A}$ and ${\cal B}$ should have the same spectrum as $A$ and $B$.

We now consider a particular case, where $A$ and $B$ are assumed to be dichotomic, $\pm 1$-valued observables, and where the approximations ${\cal A}$ and ${\cal B}$ are also constrained to be $\pm 1$-valued.
Let us recall that this same-spectrum assumption is a natural one for ${\cal B}$ in a measurement-disturbance scenario\footnote{By symmetry we also impose the same-spectrum assumption on ${\cal A}$ here; it is straightforward to adapt our study to cases where it is imposed on ${\cal B}$ only.}~\cite{ozawa2003uvr,branciard2013ete}. As mentioned before, this assumption further restricts the possible approximation strategies, and hence the allowed values of $(\epsilon_{\cal A}, \epsilon_{\cal B})$ (or of $(\epsilon_{\cal A}, \eta_{\cal B})$ in a measurement-disturbance scenario)~\cite{branciard2013ete}.
We shall indeed derive below stronger error-trade-off (or error-disturbance) relations for $\pm 1$-valued observables, when the same-spectrum assumption is imposed.

Note that in the case considered here, where $A^2 = {\cal A}^2 = B^2 = {\cal B}^2 = \one_{\cal H}$, one has $\Delta A^2 = 1- \moy{A}^2$, $\Delta B^2 = 1- \moy{B}^2$, $\Delta {\cal A}^2 = 1- \moy{{\cal A}}^2$, $\Delta {\cal B}^2 = 1- \moy{{\cal B}}^2$ (with $|\moy{A}|, |\moy{B}|, |\moy{{\cal A}}|, |\moy{{\cal B}}| \leq 1$), and Eqs.~(\ref{eq:bnds_eps_delta_Deltas_A}--\ref{eq:bnds_eps_delta_Deltas_B}) imply
\ba
\quad && \hspace{-1.3cm} 1 - \moy{A} \moy{{\cal A}} - \Delta A \sqrt{1 - \moy{{\cal A}}^2} \ \leq \ \epsilon_{\cal A}^2/2 \nonumber \\
&& \hspace{.5cm} \leq \ 1 - \moy{A} \moy{{\cal A}} + \Delta A \sqrt{1 - \moy{{\cal A}}^2} \ , \label{eq:bnds_eps_delta_Deltas_A_same_spec} \\[2mm]
\quad && \hspace{-1.3cm} 1 - \moy{B} \moy{{\cal B}} - \Delta B \sqrt{1 - \moy{{\cal B}}^2} \ \leq \ \epsilon_{\cal B}^2/2 \nonumber \\
&& \hspace{.5cm} \leq \ 1 - \moy{B} \moy{{\cal B}} + \Delta B \sqrt{1 - \moy{{\cal B}}^2} \ . \label{eq:bnds_eps_delta_Deltas_B_same_spec}
\ea
Moreover, the lower bounds above are further lower-bounded by 0 (attained when $\moy{{\cal A}} = \moy{A}$ and $\moy{{\cal B}} = \moy{B}$, resp.), while the upper bounds are further upper-bounded by 2 (attained when $\moy{{\cal A}} = -\moy{A}$ and $\moy{{\cal B}} = -\moy{B}$, resp.)---which implies in particular that there is here an absolute upper-bound of 2 on $\epsilon_{\cal A}$ and $\epsilon_{\cal B}$.
Note also that just like $\Delta A$ and $\Delta B$, $\moy{A}$ and $\moy{B}$ are fixed parameters of our problem; the variable parameters in Eqs.~(\ref{eq:bnds_eps_delta_Deltas_A_same_spec}--\ref{eq:bnds_eps_delta_Deltas_B_same_spec}), which depend on the particular approximations ${\cal A}$ and ${\cal B}$, are $\moy{{\cal A}}$, $\epsilon_{\cal A}$, $\moy{{\cal B}}$ and $\epsilon_{\cal B}$ (see the discussion in Subsection~\ref{subsec_fixed_vs_variable}).

\subsubsection{A relation for $\pm 1$-valued observables $A, B, {\cal A}$ and ${\cal B}$, $ $ and for specified values of $\moy{{\cal A}}$ and $\moy{{\cal B}}$}

Assuming $\Delta {\cal A}, \Delta {\cal B} > 0$---i.e., $|\moy{{\cal A}}|, |\moy{{\cal B}}| < 1$---one has
\ba
\frac{\Delta A^2 + \Delta {\cal A}^2 - (\epsilon_{\cal A}^2 - \delta_{\cal A}^2)}{2 \, \Delta A \, \Delta {\cal A}} & = & \frac{1 - \moy{A}\moy{{\cal A}} - \epsilon_{\cal A}^2/2}{\Delta A \, \sqrt{1- \moy{{\cal A}}^2}}, \quad \\
\frac{\Delta B^2 + \Delta {\cal B}^2 - (\epsilon_{\cal B}^2 - \delta_{\cal B}^2)}{2 \, \Delta B \, \Delta {\cal B}} & = & \frac{1 - \moy{B}\moy{{\cal B}} - \epsilon_{\cal B}^2/2}{\Delta B \, \sqrt{1- \moy{{\cal B}}^2}}, \quad
\ea
and $f_{\Delta {\cal A}, \delta_{\cal A}}(\epsilon_{\cal A}) = f_{\moy{\cal A}}(\epsilon_{\cal A})$, $f_{\Delta {\cal B}, \delta_{\cal B}}(\epsilon_{\cal B}) = f_{\moy{\cal B}}(\epsilon_{\cal B})$, with
\ba
f_{\moy{\cal A}}(\epsilon_{\cal A}) &=& \sqrt{1{-}\Big( \frac{1 - \moy{A}\moy{{\cal A}} - \epsilon_{\cal A}^2/2}{\Delta A \, \sqrt{1- \moy{{\cal A}}^2}} \Big)^{\!2} } . \label{def_fA_bis} \\
f_{\moy{\cal B}}(\epsilon_{\cal B}) &=& \sqrt{1{-}\Big( \frac{1 - \moy{B}\moy{{\cal B}} - \epsilon_{\cal B}^2/2}{\Delta B \, \sqrt{1- \moy{{\cal B}}^2}} \Big)^{\!2} } . \label{def_fB_bis}
\ea
(Note that Eqs.~(\ref{eq:bnds_eps_delta_Deltas_A_same_spec}--\ref{eq:bnds_eps_delta_Deltas_B_same_spec}) ensure that the square roots above are real.)

Thus, our relation~\eqref{eq:relation_f} implies here that when $A$ and $B$ are $\pm 1$-valued observables, and ${\cal A}$ and ${\cal B}$ are restricted to have the same spectrum ($\pm 1$) as $A$ and $B$, the possible values of $(\epsilon_{\cal A}, \epsilon_{\cal B})$ are bound to satisfy, for any specified values of $|\moy{{\cal A}}|, |\moy{{\cal B}}| < 1$,
\ba
& \hspace{-4cm} f_{\moy{\cal A}}(\epsilon_{\cal A})^2 + f_{\moy{\cal B}}(\epsilon_{\cal B})^2 \nonumber \\
&+ 2 \sqrt{1 - \tilde C_{\!AB}^2} \ f_{\moy{\cal A}}(\epsilon_{\cal A}) \, f_{\moy{\cal B}}(\epsilon_{\cal B}) \ \geq \ \tilde C_{\!AB}^2 . \quad \label{eq:relation_fbis}
\ea
This relation lower-bounds the possible values of $(f_{\moy{\cal A}}(\epsilon_{\cal A}), f_{\moy{\cal B}}(\epsilon_{\cal B}))$. As $f_{\moy{\cal A}}(\epsilon_{\cal A})$ increases with $\epsilon_{\cal A}$ for $\epsilon_{\cal A}^2/2 \leq 1{-}\moy{A}\moy{{\cal A}}$ and then decreases for $\epsilon_{\cal A}^2/2 \geq 1{-}\moy{A}\moy{{\cal A}}$ (and similarly for $f_{\moy{\cal B}}(\epsilon_{\cal B})$), the above relation bounds $\epsilon_{\cal A}$ and $\epsilon_{\cal B}$ from both below and above---with absolute lower- and upper-bounds given by~(\ref{eq:bnds_eps_delta_Deltas_A_same_spec}--\ref{eq:bnds_eps_delta_Deltas_B_same_spec}).

\medskip

In the case where $\moy{{\cal A}} = \pm 1$, $\epsilon_{\cal A}^2/2 = 1 \mp \moy{A}$ is fixed and $\epsilon_{\cal B}$ is then only bounded by Eq.~\eqref{eq:bnds_eps_delta_Deltas_B_same_spec}.
Similarly, in the case where $\moy{{\cal B}} = \pm 1$, $\epsilon_{\cal B}^2/2 = 1 \mp \moy{B}$ is fixed and $\epsilon_{\cal A}$ is only bounded by Eq.~\eqref{eq:bnds_eps_delta_Deltas_A_same_spec}.
These constraints indeed correspond to the limits of the constraint~\eqref{eq:relation_fbis} when $\moy{{\cal A}} \to \pm 1$ and $\moy{{\cal B}} \to \pm 1$, respectively.

\subsubsection{A relation for $\pm 1$-valued observables $A, B, {\cal A}$ and ${\cal B}$, $ $ valid for any values of $\moy{{\cal A}}$ and $\moy{{\cal B}}$}

As was done before, we can derive, from Eq.~\eqref{eq:relation_fbis}, an error-trade-off relation that does not involve $\moy{{\cal A}}$ and $\moy{{\cal B}}$.

Optimizing over $\moy{{\cal A}}$ (with $-1 < \moy{{\cal A}} < 1$), one can indeed prove that
\ba
1 - \Big( \frac{1 - \moy{A}\moy{{\cal A}} - \epsilon_{\cal A}^2/2}{\Delta A \, \sqrt{1- \moy{{\cal A}}^2}} \Big)^2 & \ \leq \ & \frac{1 - (1 - \epsilon_{\cal A}^2/2)^2}{\Delta A^2}  \quad
\ea 
so that for all $\moy{{\cal A}} \neq \pm 1$ and $\epsilon_{\cal A}$ satisfying~\eqref{eq:bnds_eps_delta_Deltas_A_same_spec},
\ba
&& f_{\moy{\cal A}}(\epsilon_{\cal A}) \ \leq \ k_A(\epsilon_{\cal A}) , \quad \text{with} \label{fA_leq_kA} \\[1mm]
&& k_A(\epsilon_{\cal A}) = \frac{\sqrt{1 - (1 - \epsilon_{\cal A}^2/2)^2}}{\Delta A} , \qquad \label{def_kA}
\ea
with equality in~\eqref{fA_leq_kA} if $|\moy{A}| < |1 - \epsilon_{\cal A}^2/2|$ and $\moy{{\cal A}} = \frac{\moy{A}}{1 - \epsilon_{\cal A}^2/2}$.
Contrary to $f_{\moy{\cal A}}(\epsilon_{\cal A})$, $k_A(\epsilon_{\cal A})$ is well defined for $\moy{{\cal A}} = \pm 1$, in which case, as $\epsilon_{\cal A}^2/2 = 1 \mp \moy{A}$, it takes the value 1.

Similarly, one finds that for all $\moy{{\cal B}} \neq \pm 1$ and $\epsilon_{\cal B}$ satisfying~\eqref{eq:bnds_eps_delta_Deltas_B_same_spec},
\ba
&& f_{\moy{\cal B}}(\epsilon_{\cal B}) \ \leq \ k_B(\epsilon_{\cal B}) , \quad \text{with} \label{fB_leq_kB} \\[1mm]
&& k_B(\epsilon_{\cal B}) = \frac{\sqrt{1 - (1 - \epsilon_{\cal B}^2/2)^2}}{\Delta B} , \qquad \label{def_kB}
\ea
with equality in~\eqref{fB_leq_kB} if $|\moy{B}| < |1 - \epsilon_{\cal B}^2/2|$ and $\moy{{\cal B}} = \frac{\moy{B}}{1 - \epsilon_{\cal B}^2/2}$.
Again, $k_B(\epsilon_{\cal B})$ is also well defined for $\moy{{\cal B}} = \pm 1$, in which case it takes the value 1.

Using Eqs.~\eqref{fA_leq_kA} and~\eqref{fB_leq_kB}, it follows from our previous relation~\eqref{eq:relation_fbis} that when $A$ and $B$ are $\pm 1$-valued observables, and ${\cal A}$ and ${\cal B}$ are restricted to have the same spectrum ($\pm 1$) as $A$ and $B$, the following relation must be satisfied:
\ba
k_A(\epsilon_{\cal A})^2 + k_B(\epsilon_{\cal B})^2 + 2 \sqrt{1 - \tilde C_{\!AB}^2} \ k_A(\epsilon_{\cal A}) \, k_B(\epsilon_{\cal B}) \ \geq \ \tilde C_{\!AB}^2 \nonumber \\ \label{eq:relation_PRL}
\ea
(which, strictly speaking, has so far been proven in the case where $|\moy{{\cal A}}|, |\moy{{\cal B}}| < 1$, but also trivially holds for $|\moy{{\cal A}}|, |\moy{{\cal B}}| = 1$).
This relation lower-bounds the possible values of $(k_A(\epsilon_{\cal A}), k_B(\epsilon_{\cal B}))$. As $k_A(\epsilon_{\cal A})$ increases with $\epsilon_{\cal A}$ for $\epsilon_{\cal A}^2/2 \leq 1$ and then decreases (and similarly for $k_B(\epsilon_{\cal B})$), the above relation bounds $\epsilon_{\cal A}$ and $\epsilon_{\cal B}$ from both below and above---with a trivial absolute lower-bound of 0, and an absolute upper-bound of 2 (see the discussion after Eqs.~(\ref{eq:bnds_eps_delta_Deltas_A_same_spec}--\ref{eq:bnds_eps_delta_Deltas_B_same_spec})).
Note that Eq.~\eqref{eq:relation_PRL} is strictly more restrictive than relation~\eqref{eq:relation_PNAS}, for which the same-spectrum assumption was not imposed.

Relation~\eqref{eq:relation_PRL} was introduced and experimentally tested in~\cite{Ringbauer:2013aa}, and generalizes a relation first introduced in Ref.~\cite{branciard2013ete} for the particular case where $\moy{A} = \moy{B} = 0$; our approach here allows us to prove it in a more general case, without this restriction.

\medskip

To finish this section, let us mention that the different cases and assumptions considered in the derivation of all our relations above do not need to be the same for $A, {\cal A}$ and $B, {\cal B}$. One can indeed derive ``hybrid'' relations involving for instance (say) $f_{\Delta {\cal A}, \delta_{\cal A}}(\epsilon_{\cal A})$ and $h_{\delta_{\cal B}}(\epsilon_{\cal B})$, or where the same-spectrum assumption is imposed on one observable only---which is indeed relevant in the measurement-disturbance scenario; note e.g. that the relation involving $\tilde \epsilon_{\cal A}$ and $k_B(\epsilon_{\cal B})$ has also been considered and tested experimentally in Ref.~\cite{Ringbauer:2013aa}.

\section{An alternative form for our error-trade-off relations}
\label{sec_altern_form}

Another, equivalent form for the error-trade-off relations derived in the previous section can be given, which may be convenient to use in certain cases. In this alternative form, the parameter $C_{\!AB}$ (or $\tilde C_{\!AB}$) related to the commutator of $A$ and $B$ only appears in the right-hand side, as a lower bound---as it is the case for instance in Robertson's relation~\eqref{eq:robertson}~\cite{robertson1929tup}, or in the error-trade-off relations previously derived by Ozawa~\cite{ozawa2003uvr,ozawa2004urj}, Hall~\cite{hall2004pih} and Weston \emph{et al.}~\cite{Weston:2013fk} (Eqs.~\eqref{eq_ozawa}, \eqref{eq_hall} and~\eqref{eq_weston} in Section~\ref{sec_previous_relations_follow} below).

Note first that all our above relations~\eqref{eq:relation_f}, \eqref{eq:relation_g}, \eqref{eq:relation_h}, \eqref{eq:relation_PNAS}, \eqref{eq:relation_fbis} and~\eqref{eq:relation_PRL} are of the general form
\ba
u_{\cal A}^2 + u_{\cal B}^2 + 2 \sqrt{1 - \tilde C_{\!AB}^2} \ u_{\cal A} \, u_{\cal B} \ \geq \ \tilde C_{\!AB}^2, \label{eq:relation_general_form}
\ea
with $u_{\cal A}, u_{\cal B} \geq 0$ (and $\tilde C_{\!AB}^2 \in [0,1]$).
As we show below, this turns out to be equivalent to\footnote{Note that one can similarly prove that~\eqref{eq:relation_general_form} and~\eqref{eq:relation_general_form_altern} are also equivalent to
\ba
\left\{ \!\!
\begin{array}{ll}
& u_{\cal A}^2 + u_{\cal B}^2 \geq \tilde C_{\!AB}^2 \\
\text{or} \!\! \\
& u_{\cal A}^2 + u_{\cal B}^2 \leq \tilde C_{\!AB}^2 \text{ and } u_{\cal A} \sqrt{1-u_{\cal B}^2} + u_{\cal B} \sqrt{1-u_{\cal A}^2} \, \geq \, |\tilde C_{\!AB}|.
\end{array}
\right. \nonumber
\ea
(And that the case of equality in~\eqref{eq:relation_general_form} is also equivalent to $\big[ u_{\cal A}^2 + u_{\cal B}^2 \leq \tilde C_{\!AB}^2 $ and $u_{\cal A} \sqrt{1{-}u_{\cal B}^2} + u_{\cal B} \sqrt{1{-}u_{\cal A}^2} = |\tilde C_{\!AB}| \big]$.)}
\ba
\left\{ \!\!
\begin{array}{ll}
& u_{\cal A}^2 + u_{\cal B}^2 \geq 1 \\[1mm]
\text{or} \\[1mm]
& u_{\cal A}^2 + u_{\cal B}^2 \leq 1 \text{ and } \\[1mm]
& \quad  \ u_{\cal A} \sqrt{1-u_{\cal B}^2} + u_{\cal B} \sqrt{1-u_{\cal A}^2} \, \geq \, |\tilde C_{\!AB}| .
\end{array}
\right. \quad \label{eq:relation_general_form_altern}
\ea
(Furthermore, the case where Eq.~\eqref{eq:relation_general_form} is saturated is equivalent to $\big[ u_{\cal A}^2 + u_{\cal B}^2 \leq 1 $ and $u_{\cal A} \sqrt{1{-}u_{\cal B}^2} + u_{\cal B} \sqrt{1{-}u_{\cal A}^2} = |\tilde C_{\!AB}| \big]$.)

\begin{proof}

Consider the following 3 cases:

\begin{itemize}

\item If $u_{\cal A}^2 + u_{\cal B}^2 \geq 1$, then both Eqs.~\eqref{eq:relation_general_form} and~\eqref{eq:relation_general_form_altern} trivially hold.

\item If $\tilde C_{\!AB}^2 \leq u_{\cal A}^2 + u_{\cal B}^2 \leq 1$, then Eq.~\eqref{eq:relation_general_form} still trivially holds; on the other hand, note that for $u_{\cal A}^2 + u_{\cal B}^2 \leq 1$ (which implies in particular that both $u_{\cal A}^2 \leq 1$ and $u_{\cal B}^2 \leq 1$), one has $\sqrt{1{-}u_{\cal A}^2} \, \sqrt{1{-}u_{\cal B}^2} \geq u_{\cal A} u_{\cal B}$ and hence
\ba
&& \Big( u_{\cal A} \sqrt{1{-}u_{\cal B}^2} + u_{\cal B} \sqrt{1{-}u_{\cal A}^2} \, \Big)^2 \nonumber \\
&& \quad = u_{\cal A}^2 + u_{\cal B}^2 + 2 \, u_{\cal A} \, u_{\cal B} \, \Big( \sqrt{1{-}u_{\cal A}^2} \, \sqrt{1{-}u_{\cal B}^2} - u_{\cal A} u_{\cal B} \Big) \nonumber \\
&& \quad \quad \geq \ u_{\cal A}^2 + u_{\cal B}^2 \ \geq \ \tilde C_{\!AB}^2. \nonumber
\ea
Taking the square root leads to the last inequality in~\eqref{eq:relation_general_form_altern}.

\item if $u_{\cal A}^2 + u_{\cal B}^2 < \tilde C_{\!AB}^2$, then Eq.~\eqref{eq:relation_general_form} is equivalent to
\ba
\big( \tilde C_{\!AB}^2 - u_{\cal A}^2 - u_{\cal B}^2 \big)^2 - 4 \, \big( 1 - \tilde C_{\!AB}^2 \big) \, u_{\cal A}^2 \, u_{\cal B}^2 \ \leq \ 0, \nonumber
\ea
which can be written as
\ba
&& \Big[ \tilde C_{\!AB}^2 - \Big( u_{\cal A} \sqrt{1{-}u_{\cal B}^2} - u_{\cal B} \sqrt{1{-}u_{\cal A}^2} \, \Big)^2 \Big] \nonumber \\
&& \times \Big[ \tilde C_{\!AB}^2 - \Big( u_{\cal A} \sqrt{1{-}u_{\cal B}^2} + u_{\cal B} \sqrt{1{-}u_{\cal A}^2} \, \Big)^2 \Big] \ \leq \ 0 . \nonumber
\ea
Noting that
\ba
&& \Big( u_{\cal A} \sqrt{1{-}u_{\cal B}^2} - u_{\cal B} \sqrt{1{-}u_{\cal A}^2} \, \Big)^2 \nonumber \\
&& \quad = u_{\cal A}^2 + u_{\cal B}^2 - 2 \, u_{\cal A} \, u_{\cal B} \, \Big( u_{\cal A} u_{\cal B} + \sqrt{1{-}u_{\cal A}^2} \, \sqrt{1{-}u_{\cal B}^2} \, \Big) \nonumber \\
&& \quad \quad \leq \ u_{\cal A}^2 + u_{\cal B}^2 \ < \ \tilde C_{\!AB}^2, \nonumber
\ea
we conclude from the previous equation that $C_{\!AB}^2 \leq \big( u_{\cal A} \sqrt{1{-}u_{\cal B}^2} + u_{\cal B} \sqrt{1{-}u_{\cal A}^2} \, \big)^2$, the square root of which gives again the last inequality in~\eqref{eq:relation_general_form_altern}.

\end{itemize}

Taken together, the study of the 3 cases above shows that Eqs.~\eqref{eq:relation_general_form} and~\eqref{eq:relation_general_form_altern} are indeed equivalent. (The equality case in Eq.~\eqref{eq:relation_general_form} can be analysed along similar lines as above.)
\end{proof}

The equivalence between Eqs.~\eqref{eq:relation_general_form} and~\eqref{eq:relation_general_form_altern} allows one to write all the error-trade-off relations of the previous section in a different form. For instance, our general relation~\eqref{eq:relation_PNAS} is equivalent to
\ba
\left\{ \!\!
\begin{array}{ll}
& \tilde \epsilon_{\cal A}^2 + \tilde \epsilon_{\cal B}^2 \geq 1 \\[1mm]
\text{or} \\[1mm]
& \tilde \epsilon_{\cal A}^2 + \tilde \epsilon_{\cal B}^2 \leq 1 \text{ and } \\[1mm]
& \quad  \ \tilde \epsilon_{\cal A} \sqrt{1-\tilde \epsilon_{\cal B}^2} + \tilde \epsilon_{\cal B} \sqrt{1-\tilde \epsilon_{\cal A}^2} \, \geq \, |\tilde C_{\!AB}| .
\end{array}
\right. \quad \label{eq:relation_PNAS_altern}
\ea

Note that all the alternative relations thus obtained from Eqs.~\eqref{eq:relation_f}, \eqref{eq:relation_g}, \eqref{eq:relation_h}, \eqref{eq:relation_PNAS}, \eqref{eq:relation_fbis} and~\eqref{eq:relation_PRL} could also be derived from one another as in the previous section, starting from an alternative formulation of the geometric Lemma used in the proof of Eq.~\eqref{eq:relation_f}---obtained by replacing~\eqref{eq:geom_lemma_ineq} by a relation of the form of~\eqref{eq:relation_general_form_altern}---and using the fact that when $u_{\cal A}^2 + u_{\cal B}^2 \leq 1$, $u_{\cal A} \sqrt{1{-}u_{\cal B}^2} + u_{\cal B} \sqrt{1{-}u_{\cal A}^2}$ increases with both $u_{\cal A}$ and $u_{\cal B}$.

\section{Tightness of our \\ error-trade-off relations}
\label{sec_tightness}

In this section we prove the tightness of the error-trade-off relations of Section~\ref{sec_err_trade_off_relations} (only in a some particular cases for Eqs.~\eqref{eq:relation_fbis} and~\eqref{eq:relation_PRL})---i.e. we show that any values of $(\epsilon_{\cal A}, \epsilon_{\cal B})$ that saturate these relations can be obtained, for some possible choice of ${\cal A}$ and ${\cal B}$. For simplicity we present here the proofs for the case where\footnote{Recall that $\tilde A_0 = [A-\moy{A}]/\Delta A$, $\tilde B_0 = [B-\moy{B}]/\Delta B$, that $\tilde A_0 \ket{\psi}$ and $\tilde B_0 \ket{\psi}$ are unit vectors, and hence that $| \moy{\tilde A_0 \tilde B_0} | \leq 1$ in general.
As noted in~\cite{branciard2013ete}, $| \moy{\tilde A_0 \tilde B_0} | = 1$ always holds in particular for the case of pure qubit states, where ${\cal H} = \mathbb{C}^2$ is 2-dimensional.} $| \moy{\tilde A_0 \tilde B_0} | = 1$. For our first four relations, the case where $| \moy{\tilde A_0 \tilde B_0} | < 1$ can be studied along similar lines; as it involves more tedious calculations, we present it in the Appendix.

As emphasized above all these relations are of the general form~\eqref{eq:relation_general_form} (or~\eqref{eq:relation_general_form_altern}, equivalently).
Defining $\phi = \arg \, \moy{\tilde A_0 \tilde B_0} \in [-\pi,\pi]$---such that, in the case here where $| \moy{\tilde A_0 \tilde B_0} | = 1$, $\moy{\tilde A_0 \tilde B_0} = e^{i \phi}$ and $\tilde C_{\!AB} = {\mathrm{Im}} \moy{\tilde A_0 \tilde B_0} = \sin \phi$---the values of $(u_{\cal A}, u_{\cal B})$ that saturate the relation~\eqref{eq:relation_general_form} can be parametrized by
\ba
\Big( \, u_{\cal A} = \Big| \sin \Big( \frac{\varphi + \phi}{2} \Big) \Big|, \ u_{\cal B} = \Big| \sin \Big( \frac{\varphi - \phi}{2} \Big) \Big| \ \Big) , \quad \label{eq_param_ua_ub_saturation}
\ea
for a varying value of $\varphi \in \big[ {-}|\phi|, |\phi| \big]$ if $\cos \phi \geq 0$, or $\varphi \in \big[ |\phi|, 2\pi{-}|\phi| \big]$ if $\cos \phi \leq 0$.

In order to show the tightness of our error-trade-off relations, we will prove that the corresponding functions $(u_{\cal A}, u_{\cal B})$ can indeed take these values. In the case where $u_{\cal A}$ and $u_{\cal B}$ only increase with $\epsilon_{\cal A}$ and $\epsilon_{\cal B}$, the corresponding relations (Eqs.~\eqref{eq:relation_g}, \eqref{eq:relation_h} and \eqref{eq:relation_PNAS}) only lower-bound the values of $(\epsilon_{\cal A}, \epsilon_{\cal B})$, and proving that the values of Eq.~\eqref{eq_param_ua_ub_saturation} can be reached is sufficient to show that the lower-bound is tight. On the other hand, in the case where $u_{\cal A}$ and $u_{\cal B}$ increase and then decrease with $\epsilon_{\cal A}$ and $\epsilon_{\cal B}$, the corresponding relations (Eqs.~\eqref{eq:relation_f}, \eqref{eq:relation_fbis} and \eqref{eq:relation_PRL}) both lower- and upper-bound the values of $(\epsilon_{\cal A}, \epsilon_{\cal B})$; we will then verify that the desired values of $(u_{\cal A}, u_{\cal B})$ above can be obtained from both a lower and a greater values of $\epsilon_{\cal A}$ and $\epsilon_{\cal B}$, so as to show the tightness of both the lower- and the upper-bounds on $(\epsilon_{\cal A}, \epsilon_{\cal B})$.

\subsection{Parametrization of ${\cal A}$ and ${\cal B}$}

We first introduce a particular choice\footnote{Our parametrization here is somewhat clearer, but equivalent to that used in~\cite{branciard2013ete}.} for the approximations ${\cal A}$ and ${\cal B}$. It will be sufficient here to use measurements ${\cal A}$ and ${\cal B}$ acting on ${\cal H}$ only, without introducing any ancillary system.

In the case where $| \moy{\tilde A_0 \tilde B_0} | = 1$, $\tilde A_0 \ket{\psi}$ and $\tilde B_0 \ket{\psi}$ are, up to a phase, the same unit vector, orthogonal to $\ket{\psi}$.
Let us define
\ba
\ket{v_1} = \ket{\psi}, \quad \ket{v_2} = e^{i \phi/2} \tilde A_0 \ket{\psi} = e^{-i \phi/2} \tilde B_0 \ket{\psi}, \quad
\ea
so that $\{ \ket{v_1}, \ket{v_2} \}$ forms an orthonormal basis of (the 2-dimensional) $\text{Span} \{ \ket{\psi}, \tilde A_0 \ket{\psi}, \tilde B_0 \ket{\psi} \}$.

For a given $\varphi \in \big[ {-}|\phi|, |\phi| \big]$ if $\cos \phi \geq 0$ or $\varphi \in \big[ |\phi|, 2\pi{-}|\phi| \big]$ if $\cos \phi \leq 0$, and for a parameter $\theta \in \mathbb{R}$ such that $\cos \theta \sin \theta \neq 0$ we define, with the shorthand notations $c_\theta = \cos \theta$ and $s_\theta = \sin \theta$,
\ba
\ket{m_1} &=& c_\theta \, \ket{v_1} + e^{i \varphi/2} \, s_\theta \, \ket{v_2}, \label{eq_def_m1} \\
\ket{m_2} &=& s_\theta \, \ket{v_1} - e^{i \varphi/2} \, c_\theta \, \ket{v_2}, \label{eq_def_m2}
\ea
so that $\{ \ket{m_1}, \ket{m_2} \}$ also forms an orthonormal basis of $\text{Span} \{ \ket{\psi}, \tilde A_0 \ket{\psi}, \tilde B_0 \ket{\psi} \}$. If the dimension of ${\cal H}$ is larger than 2, we complete that basis with other vectors $\ket{m_{k \geq 3}}$ orthogonal to $\ket{m_1}$ and $\ket{m_2}$ (which are hence orthogonal to $\ket{\psi}$).

The basis $\{ \ket{m_k} \}$ is chosen to define the common eigenbasis of ${\cal A}$ and ${\cal B}$; denoting by $\alpha_k$ and $\beta_k$ their eigenvalues, we thus define
\ba
{\cal A} = \sum_k \alpha_k \, \ket{m_k}\!\bra{m_k}, \quad {\cal B} = \sum_k \beta_k \, \ket{m_k}\!\bra{m_k}. \label{eq_def_Aest_Best}
\ea
With these definitions, we get
\ba
\Delta {\cal A}^2 = ( \alpha_1{-}\alpha_2 )^2 \, c_\theta^2 \, s_\theta^2 \, , & \ \ &
\Delta {\cal B}^2 = ( \beta_1{-}\beta_2 )^2 \, c_\theta^2 \, s_\theta^2 \, , \quad \ \label{eq_DAest_DBest} \\[1mm]
\delta_{\cal A} = \alpha_1 \, c_\theta^2 + \alpha_2 \, s_\theta^2 - \moy{A} \, , & \ \ &
\delta_{\cal B} = \beta_1 \, c_\theta^2 + \beta_2 \, s_\theta^2 - \moy{B} , \qquad \ \label{eq_dAest_dBest}
\ea
and
\ba
{\mathrm{Re}} \, \moy{\tilde A_0 {\cal A}} &=& \sum \alpha_k \, {\mathrm{Re}} \, \sandwich{\psi}{\tilde A_0}{m_k}\braket{m_k}{\psi} \nonumber \\[-1mm]
&=& ( \alpha_1{-}\alpha_2 ) \, c_\theta \, s_\theta \, \cos \Big( \frac{\varphi + \phi}{2} \Big) , \\[1mm]
{\mathrm{Re}} \, \moy{\tilde B_0 {\cal B}} &=& ( \beta_1{-}\beta_2 ) \, c_\theta \, s_\theta \, \cos \Big( \frac{\varphi - \phi}{2} \Big) ,
\ea
so that Eqs.~(\ref{eq:relation_eps_delta_Deltas_A}--\ref{eq:relation_eps_delta_Deltas_B}) give
\ba
\epsilon_{\cal A}^2 - \delta_{\cal A}^2 &=& \Delta A^2 + \Delta {\cal A}^2 \nonumber \\[-1mm]
&& \ - 2 \, \Delta A \, ( \alpha_1{-}\alpha_2 ) \, c_\theta \, s_\theta \, \cos \Big( \frac{\varphi + \phi}{2} \Big) , \quad \label{eq_eps_delta_A_param} \\[1mm]
\epsilon_{\cal B}^2 - \delta_{\cal B}^2 &=& \Delta B^2 + \Delta {\cal B}^2 \nonumber \\[-1mm]
&& \ - 2 \, \Delta B \, ( \beta_1{-}\beta_2 ) \, c_\theta \, s_\theta \, \cos \Big( \frac{\varphi - \phi}{2} \Big) . \qquad \label{eq_eps_delta_B_param}
\ea

\subsection{Tightness of relation~\eqref{eq:relation_f}}
\label{subsec_tightness_f}

Relation~\eqref{eq:relation_f} bounds the possible values of $(\epsilon_{\cal A}, \epsilon_{\cal B})$ when $\Delta {\cal A}, \delta_{\cal A}$, $\Delta {\cal B}$ and $\delta_{\cal B}$ are specified. Let us define, for these specified values and for some choice of $\tau_\alpha = \pm 1$ and $\tau_\beta = \pm 1$, the eigenvalues of ${\cal A}$ and ${\cal B}$ (corresponding to the eigenstates $\ket{m_k}$ parametrized above) to be
\ba
\alpha_1 &=& \moy{A} + \delta_{{\cal A}} + \tau_\alpha \, \frac{s_\theta}{c_\theta} \, \Delta {\cal A}, \label{eq_alpha1_for_f} \\
\alpha_2 &=& \moy{A} + \delta_{{\cal A}} - \tau_\alpha \, \frac{c_\theta}{s_\theta} \, \Delta {\cal A}, \label{eq_alpha2_for_f} \\
\beta_1 &=& \moy{B} + \delta_{{\cal B}} + \tau_\beta \, \frac{s_\theta}{c_\theta} \, \Delta {\cal B}, \label{eq_beta1_for_f} \\
\beta_2 &=& \moy{B} + \delta_{{\cal B}} - \tau_\beta \, \frac{c_\theta}{s_\theta} \, \Delta {\cal B}. \label{eq_beta2_for_f}
\ea
With these definitions, ${\cal A}$ and ${\cal B}$ obtained as in~\eqref{eq_def_Aest_Best} indeed give the desired values of $\Delta {\cal A}, \delta_{\cal A}$, $\Delta {\cal B}$ and $\delta_{\cal B}$ (see Eqs.~(\ref{eq_DAest_DBest}--\ref{eq_dAest_dBest})).

From Eqs.~(\ref{eq_eps_delta_A_param}--\ref{eq_eps_delta_B_param}) we then have, for $\Delta {\cal A}, \Delta {\cal B} > 0$,
\ba
\frac{\Delta A^2 + \Delta {\cal A}^2 - (\epsilon_{\cal A}^2 - \delta_{\cal A}^2)}{2 \, \Delta A \, \Delta {\cal A}} &=& \tau_\alpha \, \cos \Big( \frac{\varphi + \phi}{2} \Big), \label{eq_frac_A} \\
\frac{\Delta B^2 + \Delta {\cal B}^2 - (\epsilon_{\cal B}^2 - \delta_{\cal B}^2)}{2 \, \Delta B \, \Delta {\cal B}} &=& \tau_\beta \, \cos \Big( \frac{\varphi - \phi}{2} \Big) \label{eq_frac_B}
\ea
(independently of $\theta$),
and from the definitions~(\ref{def_fA}--\ref{def_fB}) we get, as in~\eqref{eq_param_ua_ub_saturation},
\ba
f_{\Delta {\cal A}, \delta_{\cal A}}(\epsilon_{\cal A}) & = & \Big| \sin \Big( \frac{\varphi + \phi}{2} \Big) \Big| , \label{eq_fA_sinP} \\
f_{\Delta {\cal B}, \delta_{\cal B}}(\epsilon_{\cal B}) & = & \Big| \sin \Big( \frac{\varphi - \phi}{2} \Big) \Big| , \label{eq_fB_sinM}
\ea
which saturate relation~\eqref{eq:relation_f}.
From Eqs.~(\ref{eq_frac_A}--\ref{eq_frac_B}) it appears clearly that the choice of $\tau_\alpha, \tau_\beta = \pm 1$ allows one to saturate either the lower- (for $\tau_\alpha = + \text{sign} \big[ \cos \big( \frac{\varphi + \phi}{2} \big) \big]$, $\tau_\beta = + \text{sign} \big[ \cos \big( \frac{\varphi - \phi}{2} \big) \big]$) or the upper- (for $\tau_\alpha = - \text{sign} \big[ \cos \big( \frac{\varphi + \phi}{2} \big) \big]$, $\tau_\beta = - \text{sign} \big[ \cos \big( \frac{\varphi - \phi}{2} \big) \big]$) bounds on $\epsilon_{\cal A}$ and $\epsilon_{\cal B}$. This proves the tightness of both the lower- and upper-bounds imposed by relation~\eqref{eq:relation_f}, for any specified values of $\Delta {\cal A}, \Delta {\cal B} > 0$ and of $\delta_{\cal A}, \delta_{\cal B} \in \mathbb{R}$.

Note that in the case where $\Delta {\cal A} = 0$ (resp. $\Delta {\cal B} = 0$), the choice $\varphi = \phi$ (resp. $\varphi = -\phi$) in the parametrisation of $\{\ket{m_k}\}$, together with the choice of eigenvalues given in~(\ref{eq_alpha1_for_f}--\ref{eq_beta2_for_f}), also allows one to saturate the constraints of Eq.~\eqref{eq:relation_f_limit_DA_0} (resp.~\eqref{eq:relation_f_limit_DB_0}).

\subsection{Tightness of relation~\eqref{eq:relation_g}}
\label{subsec_tightness_g}

Relation~\eqref{eq:relation_g} considers the case where $\Delta {\cal A}$ and $\Delta {\cal B}$ are specified.
Let us now define $\tau_\alpha = \text{sign} \big[ \cos \big( \frac{\varphi + \phi}{2} \big) \big]$, $\tau_\beta = \text{sign} \big[ \cos \big( \frac{\varphi - \phi}{2} \big) \big]$ and, for the specified values of $\Delta {\cal A}$ and $\Delta {\cal B}$,
\ba
\alpha_1 = \moy{A} + \tau_\alpha \, \frac{s_\theta}{c_\theta} \, \Delta {\cal A}, \label{eq_alpha1_for_g} & \quad &
\alpha_2 = \moy{A} - \tau_\alpha \, \frac{c_\theta}{s_\theta} \, \Delta {\cal A}, \label{eq_alpha2_for_g} \\
\beta_1 = \moy{B} + \tau_\beta \, \frac{s_\theta}{c_\theta} \, \Delta {\cal B}, \label{eq_beta1_for_g} & &
\beta_2 = \moy{B} - \tau_\beta \, \frac{c_\theta}{s_\theta} \, \Delta {\cal B}. \label{eq_beta2_for_g} \quad
\ea
That is, we just set $\delta_{{\cal A}}$ and $\delta_{{\cal B}}$ to 0 (which is their optimal value here) and $\tau_\alpha, \tau_\beta$ to $\text{sign} \big[ \cos \big( \frac{\varphi \pm \phi}{2} \big) \big]$ in the previous definitions~(\ref{eq_alpha1_for_f}--\ref{eq_beta2_for_f}). With these definitions, ${\cal A}$ and ${\cal B}$ indeed give the desired values of $\Delta {\cal A}$ and $\Delta {\cal B}$.

For this choice of eigenvalues, Eqs.~(\ref{eq_frac_A}--\ref{eq_frac_B}) become, in the case $\Delta {\cal A}, \Delta {\cal B} > 0$:
\ba
\frac{\Delta A^2 + \Delta {\cal A}^2 - \epsilon_{\cal A}^2}{2 \, \Delta A \, \Delta {\cal A}} &=& \Big| \cos \Big( \frac{\varphi + \phi}{2} \Big) \Big|, \label{eq_frac_A_g} \\
\frac{\Delta B^2 + \Delta {\cal B}^2 - \epsilon_{\cal B}^2}{2 \, \Delta B \, \Delta {\cal B}} &=& \Big| \cos \Big( \frac{\varphi - \phi}{2} \Big) \Big| \label{eq_frac_B_g}
\ea
(again, independently of $\theta$),
and from the definitions~\eqref{def_gA} and~\eqref{def_gB} we get, again as in~\eqref{eq_param_ua_ub_saturation},
\ba
g_{\Delta {\cal A}}(\epsilon_{\cal A}) = \Big| \sin \Big( \frac{\varphi + \phi}{2} \Big) \Big| , &\quad&
g_{\Delta {\cal B}}(\epsilon_{\cal B}) = \Big| \sin \Big( \frac{\varphi - \phi}{2} \Big) \Big| , \nonumber \\ \label{eq_gA_gB_sinP_sinM}
\ea
which saturate relation~\eqref{eq:relation_g}.
This proves the tightness of the lower-bound on $(\epsilon_{\cal A}, \epsilon_{\cal B})$ imposed by relation~\eqref{eq:relation_g}, for any specified values of $\Delta {\cal A}, \Delta {\cal B} > 0$. (Recall that relation~\eqref{eq:relation_g} does not impose any upper-bound on $\epsilon_{\cal A}$ and $\epsilon_{\cal B}$.)

Note that in the case where $\Delta {\cal A} = 0$ or $\Delta {\cal B} = 0$, the choice $\varphi = \phi$ or $-\phi$, together with the definitions~(\ref{eq_alpha1_for_g}--\ref{eq_beta2_for_g}), also allows one to saturate the constraints of Eq.~\eqref{eq:relation_g_limit_DA_0_DB_0}.

\subsection{Tightness of relation~\eqref{eq:relation_h}}
\label{subsec_tightness_h}

Instead of specifying $\Delta {\cal A}$ and $\Delta {\cal B}$, relation~\eqref{eq:relation_h} considers the case where $\delta_{\cal A}$ and $\delta_{\cal B}$ are specified.
For these given values, let us now choose
\ba
\alpha_1 &=& \moy{A} + \delta_{{\cal A}} + \Delta A \, \frac{s_\theta}{c_\theta} \, \cos \Big( \frac{\varphi + \phi}{2} \Big) , \label{eq_alpha1_for_h} \\
\alpha_2 &=& \moy{A} + \delta_{{\cal A}} - \Delta A \, \frac{c_\theta}{s_\theta} \, \cos \Big( \frac{\varphi + \phi}{2} \Big), \label{eq_alpha2_for_h} \\
\beta_1 &=& \moy{B} + \delta_{{\cal B}} + \Delta B \, \frac{s_\theta}{c_\theta} \, \cos \Big( \frac{\varphi - \phi}{2} \Big), \label{eq_beta1_for_h} \\
\beta_2 &=& \moy{B} + \delta_{{\cal B}} - \Delta B \, \frac{c_\theta}{s_\theta} \, \cos \Big( \frac{\varphi - \phi}{2} \Big). \label{eq_beta2_for_h}
\ea
That is, we set $\Delta {\cal A} = \Delta A |\cos \big( \frac{\varphi + \phi}{2} \big)|$, $\Delta {\cal B} = \Delta B |\cos \big( \frac{\varphi - \phi}{2} \big)|$ (which are their optimal values) and $\tau_\alpha = \text{sign} [\cos \big( \frac{\varphi + \phi}{2} \big)], \tau_\beta = \text{sign} [\cos \big( \frac{\varphi - \phi}{2} \big)]$ in the definitions~(\ref{eq_alpha1_for_f}--\ref{eq_beta2_for_f}). With these definitions, ${\cal A}$ and ${\cal B}$ indeed give the desired values of $\delta_{\cal A}$ and $\delta_{\cal B}$.

Using Eqs.~(\ref{eq_eps_delta_A_param}--\ref{eq_eps_delta_B_param}) we then find, for the above values,
\ba
\epsilon_{\cal A}^2 - \delta_{\cal A}^2 &=& \Delta A^2 \, \sin^2 \Big( \frac{\varphi + \phi}{2} \Big) , \label{eq_e2_d2_A} \\
\epsilon_{\cal B}^2 - \delta_{\cal B}^2 &=& \Delta B^2 \, \sin^2 \Big( \frac{\varphi - \phi}{2} \Big) \label{eq_e2_d2_B}
\ea
(independently of $\theta$),
and from the definitions~\eqref{def_hA} and~\eqref{def_hB} we get, again as in~\eqref{eq_param_ua_ub_saturation},
\ba
h_{\delta_{\cal A}}(\epsilon_{\cal A}) = \Big| \sin \Big( \frac{\varphi + \phi}{2} \Big) \Big| , &\quad&
h_{\delta_{\cal B}}(\epsilon_{\cal B}) = \Big| \sin \Big( \frac{\varphi - \phi}{2} \Big) \Big| , \nonumber \\ \label{eq_hA_hB_sinP_sinM}
\ea
which saturate relation~\eqref{eq:relation_h}.
This proves the tightness of the lower-bound on $(\epsilon_{\cal A}, \epsilon_{\cal B})$ imposed by relation~\eqref{eq:relation_h}, for any specified values of $\delta_{\cal A}, \delta_{\cal B}$. (Recall again that relation~\eqref{eq:relation_h} does not impose any upper-bound on $\epsilon_{\cal A}$ and $\epsilon_{\cal B}$.)

\subsection{Tightness of relation~\eqref{eq:relation_PNAS}}
\label{subsec_tightness_PNAS}

When no values of $\Delta {\cal A}, \delta_{\cal A}$, $\Delta {\cal B}$ or $\delta_{\cal B}$ are specified \emph{a priori}, let us define the eigenvalues
\ba
\alpha_1 &=& \moy{A} + \Delta A \, \frac{s_\theta}{c_\theta} \, \cos \Big( \frac{\varphi + \phi}{2} \Big) , \label{eq_alpha1_for_PNAS} \\
\alpha_2 &=& \moy{A} - \Delta A \, \frac{c_\theta}{s_\theta} \, \cos \Big( \frac{\varphi + \phi}{2} \Big), \label{eq_alpha2_for_PNAS} \\
\beta_1 &=& \moy{B} + \Delta B \, \frac{s_\theta}{c_\theta} \, \cos \Big( \frac{\varphi - \phi}{2} \Big), \label{eq_beta1_for_PNAS} \\
\beta_2 &=& \moy{B} - \Delta B \, \frac{c_\theta}{s_\theta} \, \cos \Big( \frac{\varphi - \phi}{2} \Big). \label{eq_beta2_for_PNAS}
\ea
That is, we just set $\delta_{{\cal A}}$ and $\delta_{{\cal B}}$ to 0 (which is again their optimal value here) in the previous definitions~(\ref{eq_alpha1_for_h}--\ref{eq_beta2_for_h}).
Eqs.~(\ref{eq_e2_d2_A}--\ref{eq_e2_d2_B}) then simply give
\ba
\tilde \epsilon_{\cal A} = \frac{\epsilon_{\cal A}}{\Delta A} = \Big| \sin \Big( \frac{\varphi + \phi}{2} \Big) \Big| , &\quad&
\tilde \epsilon_{\cal B} = \frac{\epsilon_{\cal B}}{\Delta B}  = \Big| \sin \Big( \frac{\varphi - \phi}{2} \Big) \Big| \nonumber \\ \label{eq_epsA_epsB_sinP_sinM}
\ea
(independently of $\theta$), which saturate relation~\eqref{eq:relation_PNAS}.
This proves the tightness of our general error-trade-off relation~\eqref{eq:relation_PNAS} (which had already been proven in~\cite{branciard2013ete}).

\medskip

Note that in this case, the above-chosen values of $\alpha_k$ and $\beta_k$ (Eqs.~(\ref{eq_alpha1_for_PNAS}--\ref{eq_beta2_for_PNAS})) are equal to $\alpha_k = \moy{A} + \Delta A \, {\mathrm{Re}} \frac{\sandwich{m_k}{\tilde A_0}{\psi}}{\braket{m_k}{\psi}} = {\mathrm{Re}} \frac{\sandwich{m_k}{A}{\psi}}{\braket{m_k}{\psi}}$ and $\beta_k = \moy{B} + \Delta B \, {\mathrm{Re}} \frac{\sandwich{m_k}{\tilde B_0}{\psi}}{\braket{m_k}{\psi}} = {\mathrm{Re}} \frac{\sandwich{m_k}{B}{\psi}}{\braket{m_k}{\psi}}$. These correspond to the ``weak values'' of $A$ and $B$ in the preselected state $\ket{\psi}$ and the postselected state $\ket{m_k}$~\cite{Aharonov:1988zr}, which are indeed optimal to minimize, for a given projection eigenbasis $\{\ket{m_k}\}$, the root-mean-square errors $\epsilon_{\cal A}, \epsilon_{\cal B}$~\cite{hall2004pih,branciard2013ete}.
Furthermore, these optimal choices for the approximate observables ${\cal A}$ and ${\cal B}$ satisfy $\Delta {{\cal A}}^2 = \Delta A^2 - \epsilon_{\cal A}^2$ and $\Delta {{\cal B}}^2 = \Delta B^2 - \epsilon_{\cal B}^2$, as noted in Ref.~\cite{hall2004pih}.

\subsection{Tightness of relations~\eqref{eq:relation_fbis} and~\eqref{eq:relation_PRL} for \newline $\moy{{\cal A}} = \moy{{\cal B}} = 0$ and $\moy{A} = \moy{B} = 0$, resp. \newline (and $|\moy{\tilde A_0 \tilde B_0}| = 1$) $\quad $}

We now consider the case where $A$, $B$, ${\cal A}$ and ${\cal B}$ are all $\pm 1$-valued observables, which are the assumptions under which relations~\eqref{eq:relation_fbis} and~\eqref{eq:relation_PRL} hold.

Note that our choice of eigenbasis $\{\ket{m_k}\}$ introduced above allows for only two possible projection outcomes when measuring $\ket{\psi}$: $\ket{m_1}$ or $\ket{m_2}$. As the corresponding eigenvalues of ${\cal A}$ and ${\cal B}$ are assumed here to be $\pm 1$, this forces the values of $\moy{{\cal A}}$ and $\moy{{\cal B}}$, with our choice of $\{\ket{m_k}\}$, to satisfy either $|\moy{{\cal A}}| = 1$, $|\moy{{\cal B}}| = 1$, or $|\moy{{\cal A}}| = |\moy{{\cal B}}| = \big| |\braket{m_1}{\psi}|^2 - |\braket{m_2}{\psi}|^2 \big| = |c_\theta^2 - s_\theta^2|$. This restricts the cases that our construction allows us to consider. In fact, we shall now prove the tightness of relations~\eqref{eq:relation_fbis} and~\eqref{eq:relation_PRL} in even further restricted cases, namely when $\moy{{\cal A}} = \moy{{\cal B}} = 0$ for relation~\eqref{eq:relation_fbis}, and when $\moy{A} = \moy{B} = 0$ for relation~\eqref{eq:relation_PRL}.

Note that as before, the proof here is presented for the case $|\moy{\tilde A_0 \tilde B_0}| = 1$. Contrary to the previous relations however, we do not provide in the Appendix a similar proof for the case $|\moy{\tilde A_0 \tilde B_0}| < 1$, and leave that (together with the cases where $\moy{{\cal A}}, \moy{{\cal B}} \neq 0$ or $\moy{A}, \moy{B} \neq 0$) as an open problem.

\subsubsection{Tightness of relation~\eqref{eq:relation_fbis} for $\moy{{\cal A}} = \moy{{\cal B}} = 0$}

Let us use the value $\theta = \frac{\pi}{4}$ in the definition of our projection eigenbasis $\{\ket{m_k}\}$ above,
and let us define the eigenvalues of ${\cal A}$ and ${\cal B}$ to be
\ba
\alpha_1 = - \alpha_2 = \tau_\alpha = \pm 1, \quad \beta_1 = - \beta_2 = \tau_\beta = \pm 1. \quad \label{eq_alphas_betas_for_same_spec}
\ea
These give the desired average values $\moy{{\cal A}} = \moy{{\cal B}} = 0$ (and $\Delta {\cal A} = \Delta {\cal B} = 1$).
Using Eqs~(\ref{eq_eps_delta_A_param}--\ref{eq_eps_delta_B_param}) (with $\Delta A^2 = 1 - \moy{A}^2$, $\Delta B^2 = 1 - \moy{B}^2$ and $c_\theta \, s_\theta = \frac{1}{2}$) we then find
\ba
\epsilon_{\cal A}^2/2 &=& 1 - \tau_\alpha \, \Delta A \, \cos \Big( \frac{\varphi + \phi}{2} \Big) , \label{eq_epsA_same_spec_moyAest0} \\
\epsilon_{\cal B}^2/2 &=& 1 - \tau_\beta \, \Delta B \, \cos \Big( \frac{\varphi - \phi}{2} \Big) , \quad  \label{eq_epsB_same_spec_moyBest0}
\ea
and from Eqs.~(\ref{def_fA_bis}--\ref{def_fB_bis}) we get, again as in~\eqref{eq_param_ua_ub_saturation},
\ba
f_{\moy{{\cal A}}=0}(\epsilon_{\cal A}) = \sqrt{1{-}\Big( \frac{1 - \epsilon_{\cal A}^2/2}{\Delta A} \Big)^{\!2} } & = & \Big| \sin \Big( \frac{\varphi + \phi}{2} \Big) \Big| , \label{eq_f_A0_same_spec} \\
f_{\moy{{\cal B}}=0}(\epsilon_{\cal B}) = \sqrt{1{-}\Big( \frac{1 - \epsilon_{\cal B}^2/2}{\Delta B} \Big)^{\!2} } & = & \Big| \sin \Big( \frac{\varphi - \phi}{2} \Big) \Big| , \qquad \label{eq_f_B0_same_spec}
\ea
which saturate relation~\eqref{eq:relation_fbis}.
From Eqs.~(\ref{eq_epsA_same_spec_moyAest0}--\ref{eq_epsB_same_spec_moyBest0}), it appears clearly that the choice of $\tau_\alpha, \tau_\beta = \pm \text{sign} \big[ \cos \big( \frac{\varphi \pm \phi}{2} \big) \big]$ allows one to saturate either the corresponding lower- or the corresponding upper-bounds on $\epsilon_{\cal A}$ and $\epsilon_{\cal B}$.
This proves the tightness of relation~\eqref{eq:relation_fbis}, for the case where $\moy{{\cal A}} = \moy{{\cal B}} = 0$ and $|\moy{\tilde A_0 \tilde B_0}| = 1$.
Proving its tightness in the general case\footnote{\label{footnote_tightness_open_pb} From the discussion above, this would require one to use a different projection eigenbasis from $\{\ket{m_k}\}$---one that is not restricted to the 2-dimensional $\text{Span} \{ \ket{\psi}, \tilde A_0 \ket{\psi}, \tilde B_0 \ket{\psi} \}$---and to introduce in general an ancillary system (or equivalently to consider general POVMs).}, or otherwise deriving a tighter relation, is left as an open problem.

\subsubsection{Tightness of relation~\eqref{eq:relation_PRL} for $\moy{A} = \moy{B} = 0$}

Assuming now that $\moy{A} = \moy{B} = 0$, and hence $\Delta A = \Delta B = 1$, the previous choice of $\alpha_k$ and $\beta_k$ (i.e. of ${\cal A}$ and ${\cal B}$) gives, for the definitions of Eqs.~\eqref{def_kA} and~\eqref{def_kB} (and using~(\ref{eq_epsA_same_spec_moyAest0}--\ref{eq_epsB_same_spec_moyBest0})):
\ba
k_{A, \Delta A = 1}(\epsilon_{\cal A}) = \sqrt{1{-}\big( 1 - \epsilon_{\cal A}^2/2 \big)^{\!2} } & = & \Big| \sin \Big( \frac{\varphi + \phi}{2} \Big) \Big| , \\
k_{B, \Delta B = 1}(\epsilon_{\cal B}) = \sqrt{1{-}\big( 1 - \epsilon_{\cal B}^2/2 \big)^{\!2} } & = & \Big| \sin \Big( \frac{\varphi - \phi}{2} \Big) \Big| , \qquad
\ea
which saturate relation~\eqref{eq:relation_PRL}.
Again, the choice of $\tau_\alpha, \tau_\beta = \pm \text{sign} \big[ \cos \big( \frac{\varphi \pm \phi}{2} \big) \big]$ allows one to saturate either the corresponding lower- or upper-bounds on $\epsilon_{\cal A}$ and $\epsilon_{\cal B}$.
This proves the tightness of our relation~\eqref{eq:relation_PRL}, for the case where $\moy{A} = \moy{B} = 0$ and $|\moy{\tilde A_0 \tilde B_0}| = 1$ (which had already been proven in~\cite{branciard2013ete}).
As before, proving its tightness in the general case\textsuperscript{\ref{footnote_tightness_open_pb}}, or otherwise deriving a tighter relation, is left as an open problem.

\section{Ozawa's~\cite{ozawa2004urj}, Hall's~\cite{hall2004pih} and Weston \emph{et al.}'s~\cite{Weston:2013fk} error-trade-off relations all follow from our tight relations}
\label{sec_previous_relations_follow}

Ozawa~\cite{ozawa2004urj}, Hall~\cite{hall2004pih} and Weston \emph{et al.}~\cite{Weston:2013fk} have previously also derived universally valid error-trade-off relations that restrain the possible values of $(\epsilon_{\cal A}, \epsilon_{\cal B})$ for all possible approximate joint measurements. We now review these and show explicitly that they follow from the relations we derived in Section~\ref{sec_err_trade_off_relations}---which are thus stronger than these (non-tight) previous relations.

\subsection{Ozawa's relation~\cite{ozawa2004urj}}

Ozawa's error-trade-off relation~\cite{ozawa2004urj} (first derived in terms of error-disturbance trade-offs~\cite{ozawa2003uvr}), states that
\ba
\epsilon_{\cal A} \, \epsilon_{\cal B} + \Delta B \, \epsilon_{\cal A} + \Delta A \, \epsilon_{\cal B} \ \geq \ |C_{\!AB}| , \label{eq_ozawa}
\ea
or equivalently (when $\Delta A \, \Delta B > 0$, with $\tilde \epsilon_{\cal A} = \frac{\epsilon_{\cal A}}{\Delta A}$ and $\tilde \epsilon_{\cal B} = \frac{\epsilon_{\cal B}}{\Delta B}$ as before)
\ba
\tilde \epsilon_{\cal A} \, \tilde \epsilon_{\cal B} + \tilde \epsilon_{\cal A} + \tilde \epsilon_{\cal B} \ \geq \ |\tilde C_{\!AB}| .
\ea

We already showed in Ref.~\cite{branciard2013ete} that this relation follows directly from our relation~\eqref{eq:relation_PNAS}. This can also easily be verified using the alternative (equivalent) form of Eq.~\eqref{eq:relation_PNAS_altern}:
\ba
\left\{ \!
\begin{array}{ll}
\text{if } \tilde \epsilon_{\cal A}^2 + \tilde \epsilon_{\cal B}^2 \geq 1: \\[1mm]
\quad \tilde \epsilon_{\cal A} \, \tilde \epsilon_{\cal B} \! + \! \tilde \epsilon_{\cal A} \! + \! \tilde \epsilon_{\cal B} \geq \tilde \epsilon_{\cal A} \! + \! \tilde \epsilon_{\cal B} \geq \sqrt{\epsilon_{\cal A}^2 \! + \! \tilde \epsilon_{\cal B}^2} \geq 1 \geq |\tilde C_{\!AB}| , \hspace{-5mm} \\[4mm]
\text{if } \tilde \epsilon_{\cal A}^2 + \tilde \epsilon_{\cal B}^2 \leq 1: \\[1mm]
\quad  \tilde \epsilon_{\cal A} \, \tilde \epsilon_{\cal B} + \tilde \epsilon_{\cal A} + \tilde \epsilon_{\cal B} \geq \tilde \epsilon_{\cal A} + \tilde \epsilon_{\cal B} \\
\qquad \quad \geq \tilde \epsilon_{\cal A} \sqrt{1-\tilde \epsilon_{\cal B}^2} + \tilde \epsilon_{\cal B} \sqrt{1-\tilde \epsilon_{\cal A}^2} \geq |\tilde C_{\!AB}| .
\end{array}
\right. \quad \quad
\ea

As can be seen above (and as already noted in~\cite{branciard2013ete}), Ozawa's relation actually remains valid if one drops the first product term $\tilde \epsilon_{\cal A} \, \tilde \epsilon_{\cal B}$. Furthermore, his relation (even without the first product term) can only be saturated in the very specific case where $\tilde \epsilon_{\cal A} = 0$ or $\tilde \epsilon_{\cal B} = 0$. On the contrary, as shown in the previous section our relation~\eqref{eq:relation_PNAS} (or~\eqref{eq:relation_PNAS_altern}, equivalently) can always be saturated, and is thus strictly stronger than Ozawa's.

\subsection{Hall's relation~\cite{hall2004pih}}
\label{subsec_hall}

Soon after Ozawa, Hall~\cite{hall2004pih} derived the relation
\ba
\epsilon_{\cal A} \, \epsilon_{\cal B} + \Delta {\cal B} \, \epsilon_{\cal A} + \Delta {\cal A} \, \epsilon_{\cal B} \ \geq \ |C_{\!AB}|. \label{eq_hall}
\ea
Although Hall's relation looks quite similar to Ozawa's, it is in fact fundamentally different. Note indeed that instead of the (fixed) standard deviations $\Delta A, \Delta B$ of the observables $A$ and $B$, it involves the standard deviations $\Delta {\cal A}, \Delta {\cal B}$ of the approximations ${\cal A}$ and ${\cal B}$.
In particular, contrary to Ozawa's relation, one cannot drop the first product term $\epsilon_{\cal A} \, \epsilon_{\cal B}$ in Hall's relation (when $|C_{\!AB}| > 0$), as it is for instance always possible to choose approximate observables ${\cal A}, {\cal B}$ such that $\Delta {\cal A} = \Delta {\cal B} = 0$, and hence $\Delta {\cal A} \, \epsilon_{\cal B} + \Delta {\cal B} \, \epsilon_{\cal A}$ cannot be lower-bounded by any strictly positive term.

Because Hall's relation involves the parameters $\Delta {\cal A}, \Delta {\cal B}$ that depend on the particular choice of ${\cal A}$ and ${\cal B}$, it cannot---contrary to Ozawa's---be derived from~\eqref{eq:relation_PNAS}. We show below that it follows instead from our relation~\eqref{eq:relation_g} (which precisely also involves $\Delta {\cal A}, \Delta {\cal B}$).
For that, let us first show more generally that from any relation of the form~\eqref{eq:relation_general_form}---or~\eqref{eq:relation_general_form_altern}, equivalently---with $u_{\cal A}, u_{\cal B} \leq 1$, it follows that, for any $\textsc{x}, \textsc{y} \geq 0$:
\ba
&& \sqrt{u_{\cal A}^2{+}\big( \!\sqrt{1{-}u_{\cal A}^2}{-}\textsc{x} \big)^2} \ \sqrt{u_{\cal B}^2{+}\big( \!\sqrt{1{-}u_{\cal B}^2}{-}\textsc{y} \big)^2} \nonumber \\
&& + \textsc{y} \sqrt{u_{\cal A}^2{+}\big( \!\sqrt{1{-}u_{\cal A}^2}{-}\textsc{x} \big)^2} + \textsc{x} \sqrt{u_{\cal B}^2{+}\big( \!\sqrt{1{-}u_{\cal B}^2}{-}\textsc{y} \big)^2} \geq |\tilde C_{\!AB}| . \nonumber \\ \label{eq:general_3cauchy}
\ea

\begin{proof}
Similarly to Hall's derivation~\cite{hall2004pih}, our proof is based on the decomposition of a scalar product $\hat a \cdot \hat b$ in the form $\hat a \cdot \hat b = (\hat a{-}\vec x) \cdot (\hat b{-}\vec y) + \hat a \cdot \vec y + \vec x \cdot \hat b$, where $\vec x, \vec y$ are two orthogonal vectors ($\vec x \cdot \vec y = 0$), and on the application of the Cauchy-Schwartz (CS) inequality to the three scalar products of the decomposition.
Let us indeed consider the following 2 cases:

\begin{itemize}

\item Assume first that $u_{\cal A}^2 + u_{\cal B}^2 \leq 1$. From the CS inequality, we have
\ba
&& \left(\!\! \begin{array}{c} \sqrt{1{-}u_{\cal A}^2} - \textsc{x} \\ u_{\cal A} \end{array} \!\!\right) \cdot \left(\!\! \begin{array}{c} u_{\cal B} \\ \sqrt{1{-}u_{\cal B}^2} - \textsc{y} \end{array} \!\!\right) \nonumber \\
&& \qquad \leq \sqrt{u_{\cal A}^2{+}\big( \!\sqrt{1{-}u_{\cal A}^2}{-}\textsc{x} \big)^2} \ \sqrt{u_{\cal B}^2{+}\big( \!\sqrt{1{-}u_{\cal B}^2}{-}\textsc{y} \big)^2} , \nonumber \\
&& \left(\!\! \begin{array}{c} \sqrt{1{-}u_{\cal A}^2} - \textsc{x} \\ u_{\cal A} \end{array} \!\!\right) \cdot \left(\!\! \begin{array}{c} 0 \\ \textsc{y} \end{array} \!\!\right) \ \leq \ \textsc{y} \ \sqrt{u_{\cal A}^2{+}\big( \!\sqrt{1{-}u_{\cal A}^2}{-}\textsc{x} \big)^2} , \nonumber \\
&& \left(\!\! \begin{array}{c} \textsc{x} \\ 0 \end{array} \!\!\right) \cdot \left(\!\! \begin{array}{c} u_{\cal B} \\ \sqrt{1{-}u_{\cal B}^2} - \textsc{y} \end{array} \!\!\right) \ \leq \ \textsc{x} \ \sqrt{u_{\cal B}^2{+}\big( \!\sqrt{1{-}u_{\cal B}^2}{-}\textsc{y} \big)^2} . \nonumber
\ea
By summing the right-hand side (r.h.s.) of the three inequalities above, we get the left-hand side (l.h.s.) of Eq.~\eqref{eq:general_3cauchy}. On the other hand, the sum of the l.h.s. of the three inequalities above gives
\ba
\left(\!\! \begin{array}{c} \sqrt{1{-}u_{\cal A}^2} \\ u_{\cal A} \end{array} \!\right) \cdot \left(\!\! \begin{array}{c} u_{\cal B} \\ \sqrt{1{-}u_{\cal B}^2} \end{array} \!\right) = u_{\cal A} \sqrt{1{-}u_{\cal B}^2} + u_{\cal B} \sqrt{1{-}u_{\cal A}^2} \nonumber
\ea
which, by assumption (from Eq.~\eqref{eq:relation_general_form_altern}, in the case where $u_{\cal A}^2 + u_{\cal B}^2 \leq 1$), is lower-bounded by $|\tilde C_{\!AB}|$. Summing the three inequalities above thus leads to Eq.~\eqref{eq:general_3cauchy}.

\item Assume now that $u_{\cal A}^2 + u_{\cal B}^2 \geq 1$. Following similar calculations as in the previous case, but replacing $u_{\cal B}$ by $\sqrt{1{-}u_{\cal A}^2}$, leads to a similar inequality as Eq.~\eqref{eq:general_3cauchy}, where $u_{\cal B}^2{+}\big( \!\sqrt{1{-}u_{\cal B}^2}{-}\textsc{y} \big)^2$ is replaced by $1{-}u_{\cal A}^2{+}\big( u_{\cal A}{-}\textsc{y} \big)^2$ and the r.h.s. is 1 ($\geq |\tilde C_{\!AB}|$). Now, one can easily check that when $u_{\cal A}^2 + u_{\cal B}^2 \geq 1$, $u_{\cal B}^2{+}\big( \!\sqrt{1{-}u_{\cal B}^2}{-}\textsc{y} \big)^2 \geq 1{-}u_{\cal A}^2{+}\big( u_{\cal A}{-}\textsc{y} \big)^2$, which allows one to conclude that Eq.~\eqref{eq:general_3cauchy} still holds.
\end{itemize}
\end{proof}

In the case where $\Delta {\cal A}, \Delta {\cal B} > 0$, relation~\eqref{eq:relation_g} thus implies Eq.~\eqref{eq:general_3cauchy} with $u_{\cal A} = g_{\Delta {\cal A}}(\epsilon_{\cal A})$ and $u_{\cal B} = g_{\Delta {\cal B}}(\epsilon_{\cal B})$; taking $\textsc{x} = \frac{\Delta {\cal A}}{\Delta A}$ and $\textsc{y} = \frac{\Delta {\cal B}}{\Delta B}$, this leads, after simplification and multiplication by $\Delta A \, \Delta B$, to
\ba
\epsilon_{\cal A}^{(min)} \, \epsilon_{\cal B}^{(min)} + \Delta {\cal B} \, \epsilon_{\cal A}^{(min)} + \Delta {\cal A} \, \epsilon_{\cal B}^{(min)} \ \geq \ |C_{\!AB}|, \qquad
\ea
with $\epsilon_{\cal A}^{(min)} = \min\big[ \epsilon_{\cal A}, \sqrt{\Delta A^2 + \Delta {\cal A}^2} \big]$ and $\epsilon_{\cal B}^{(min)} = \min\big[ \epsilon_{\cal B}, \sqrt{\Delta B^2 + \Delta {\cal B}^2} \big]$. Using $\epsilon_{\cal A} \geq \epsilon_{\cal A}^{(min)}$ and $\epsilon_{\cal B} \geq \epsilon_{\cal B}^{(min)}$, this in turn implies Hall's relation~\eqref{eq_hall}. In the case where $\Delta {\cal A} = 0$, using Eq.~(\ref{eq:relation_g_limit_DA_0_DB_0}, left) we directly find
\ba
\epsilon_{\cal A} \, \epsilon_{\cal B} + \Delta {\cal B} \, \epsilon_{\cal A} \ & \geq & \ \Delta A \, \big( |\Delta B - \Delta {\cal B}| + \Delta {\cal B} \big) \nonumber \\
& \geq & \ \Delta A \, \Delta B \ \geq \ |C_{\!AB}|,
\ea
which gives again Hall's relation (with $\Delta {\cal A} = 0$); the case where $\Delta {\cal B} = 0$ is obtained similarly.

The independent use of 3 CS inequalities in the proof above, which cannot in general all be saturated simultaneously, explains the non-optimality of Hall's relation, as opposed to our relation~\eqref{eq:relation_g}. Looking at the saturation conditions for the 3 CS inequalities, one can see that Hall's relation can only be saturated if $\epsilon_{\cal A} = 0$ or $\epsilon_{\cal B} = 0$, or if $\Delta {\cal A} \, \Delta {\cal B} = 0$, $\epsilon_{\cal A} = \Delta A - \Delta {\cal A}$, $\epsilon_{\cal B} = \Delta B - \Delta {\cal B}$ and $A$, $B$ and $\ket{\psi}$ saturate Robertson's relation (i.e. $\Delta A \, \Delta B = |C_{\!AB}|$, or equivalently $|\tilde C_{\!AB}| = 1$).

\medskip

Note that other (non-tight) inequalities can be derived from our relations of Section~\ref{sec_err_trade_off_relations} and by using Eq.~\eqref{eq:general_3cauchy}, for different values of $\textsc{x}$ and $\textsc{y}$. For instance, from Eq.~\eqref{eq:relation_PNAS}---more precisely, from a similar (in fact, equivalent) relation to Eq.~\eqref{eq:relation_PNAS} where $\tilde \epsilon_{\cal A}$ and $\tilde \epsilon_{\cal B}$ are replaced by $\bar \epsilon_{\cal A} = \min[\tilde \epsilon_{\cal A},1]$ and $\bar \epsilon_{\cal B} = \min[\tilde \epsilon_{\cal B},1]$, resp.---and with the choice $\textsc{x} = \sqrt{1-\bar \epsilon_{\cal A}^2}$ and $\textsc{y} = \sqrt{1-\bar \epsilon_{\cal B}^2}$, one obtains
\ba
\bar \epsilon_{\cal A} \, \bar \epsilon_{\cal B} + \sqrt{1-\bar \epsilon_{\cal B}^2} \, \bar \epsilon_{\cal A} + \sqrt{1-\bar \epsilon_{\cal A}^2} \, \bar \epsilon_{\cal B} \ \geq \ |\tilde C_{\!AB}| , \quad
\ea
which in turn implies Ozawa's relation~\eqref{eq_ozawa}.
The non-optimality of Ozawa's relation---and in particular its first, unnecessary product term---thus also appear to be due to the independent use of 3 CS inequalities in its proof~\cite{ozawa2003uvr,ozawa2004urj} which cannot in general be saturated simultaneously.

\subsection{Weston \emph{et al.}'s relation~\cite{Weston:2013fk}}
\label{subsec_weston}

A new error-trade-off relation was recently derived by Weston \emph{et al.}~\cite{Weston:2013fk}, which reads
\ba
\frac{\Delta B + \Delta {\cal B}}{2} \ \epsilon_{\cal A}  \ + \ \frac{\Delta A + \Delta {\cal A}}{2} \ \epsilon_{\cal B}  \ \geq \ |C_{\!AB}|. \quad \label{eq_weston}
\ea

We will show in a similar manner that this relation can be derived from~\eqref{eq:relation_PNAS}.
Let us here first show more generally that from any relation of the form~\eqref{eq:relation_general_form}---or~\eqref{eq:relation_general_form_altern}, equivalently---with $u_{\cal A}, u_{\cal B} \leq 1$, it follows that, for any $\textsc{x}, \textsc{y} \geq 0$:
\ba
&& \frac{1+\textsc{y}}{2} \, \sqrt{u_{\cal A}^2{+}\big( \!\sqrt{1{-}u_{\cal A}^2}{-}\textsc{x} \big)^2} \nonumber \\
&& \quad + \frac{1+\textsc{x}}{2} \, \sqrt{u_{\cal B}^2{+}\big( \!\sqrt{1{-}u_{\cal B}^2}{-}\textsc{y} \big)^2} \ \ \geq \ |\tilde C_{\!AB}| . \quad \label{eq:general_4cauchy}
\ea

\begin{proof}

The proof here uses a similar trick as the proof of Eq.~\eqref{eq:general_3cauchy} above, using now (as in Weston \emph{et al.}'s derivation~\cite{Weston:2013fk}) the decomposition of a scalar product $\hat a \cdot \hat b$ in the form $\hat a \cdot \hat b = \frac{1}{2} \big[ (\hat a{-}\vec x) \cdot \hat b + \hat a \cdot (\hat b{-}\vec y) + (\hat a{-}\vec x) \cdot \vec y + \vec x \cdot (\hat b{-}\vec y) \big]$, where $\vec x, \vec y$ are two orthogonal vectors ($\vec x \cdot \vec y = 0$), and applying the CS inequality to the four scalar products of the decomposition.
We consider again the following 2 cases:

\begin{itemize}

\item Assume first that $u_{\cal A}^2 + u_{\cal B}^2 \leq 1$. From the CS inequality, we have
\ba
\left(\!\! \begin{array}{c} \sqrt{1{-}u_{\cal A}^2} - \textsc{x} \\ u_{\cal A} \end{array} \!\!\right) \cdot \left(\!\! \begin{array}{c} u_{\cal B} \\ \sqrt{1{-}u_{\cal B}^2} \end{array} \!\right) & \leq & \sqrt{u_{\cal A}^2{+}\big( \!\sqrt{1{-}u_{\cal A}^2}{-}\textsc{x} \big)^2} , \nonumber \\
\left(\!\! \begin{array}{c} \sqrt{1{-}u_{\cal A}^2} \\ u_{\cal A} \end{array} \!\right) \cdot \left(\!\! \begin{array}{c} u_{\cal B} \\ \sqrt{1{-}u_{\cal B}^2} - \textsc{y} \end{array} \!\!\right) & \leq & \sqrt{u_{\cal B}^2{+}\big( \!\sqrt{1{-}u_{\cal B}^2}{-}\textsc{y} \big)^2} , \nonumber \\
\left(\! \begin{array}{c} \textsc{x} \\ 0 \end{array} \!\right) \cdot \left(\!\! \begin{array}{c} u_{\cal B} \\ \sqrt{1{-}u_{\cal B}^2} - \textsc{y} \end{array} \!\!\right) & \leq & \textsc{x} \, \sqrt{u_{\cal B}^2{+}\big( \!\sqrt{1{-}u_{\cal B}^2}{-}\textsc{y} \big)^2} , \nonumber \\
\left(\!\! \begin{array}{c} \sqrt{1{-}u_{\cal A}^2} - \textsc{x} \\ u_{\cal A} \end{array} \!\!\right) \cdot \left(\! \begin{array}{c} 0 \\ \textsc{y} \end{array} \!\right) & \leq & \textsc{y} \, \sqrt{u_{\cal A}^2{+}\big( \!\sqrt{1{-}u_{\cal A}^2}{-}\textsc{x} \big)^2} . \nonumber
\ea
By summing the r.h.s. of the four inequalities above and dividing by 2, we get the l.h.s. of Eq.~\eqref{eq:general_4cauchy}. On the other hand, the sum of the l.h.s. of the four inequalities above, divided by 2, gives again
\ba
\left(\!\! \begin{array}{c} \sqrt{1{-}u_{\cal A}^2} \\ u_{\cal A} \end{array} \!\right) \cdot \left(\!\! \begin{array}{c} u_{\cal B} \\ \sqrt{1{-}u_{\cal B}^2} \end{array} \!\right) = u_{\cal A} \sqrt{1{-}u_{\cal B}^2} + u_{\cal B} \sqrt{1{-}u_{\cal A}^2} \nonumber
\ea
which, by assumption (from Eq.~\eqref{eq:relation_general_form_altern}, in the case where $u_{\cal A}^2 + u_{\cal B}^2 \leq 1$), is lower-bounded by $|\tilde C_{\!AB}|$. Summing the four inequalities above thus leads to Eq.~\eqref{eq:general_4cauchy}.

\item Assume now that $u_{\cal A}^2 + u_{\cal B}^2 \geq 1$. Following again similar calculations as in the previous case, but replacing $u_{\cal B}$ by $\sqrt{1{-}u_{\cal A}^2}$, leads to a similar inequality as Eq.~\eqref{eq:general_4cauchy}, where $u_{\cal B}^2{+}\big( \!\sqrt{1{-}u_{\cal B}^2}{-}\textsc{y} \big)^2$ is replaced by $1{-}u_{\cal A}^2{+}\big( u_{\cal A}{-}\textsc{y} \big)^2$ and the r.h.s. is 1 ($\geq |\tilde C_{\!AB}|$). As previously, using the fact that when $u_{\cal A}^2 + u_{\cal B}^2 \geq 1$, $u_{\cal B}^2{+}\big( \!\sqrt{1{-}u_{\cal B}^2}{-}\textsc{y} \big)^2 \geq 1{-}u_{\cal A}^2{+}\big( u_{\cal A}{-}\textsc{y} \big)^2$ allows one to conclude that Eq.~\eqref{eq:general_4cauchy} still holds.
\end{itemize}
\end{proof}

Similarly to the previous subsection, in the case where $\Delta {\cal A}, \Delta {\cal B} > 0$, relation~\eqref{eq:relation_g} also implies Eq.~\eqref{eq:general_4cauchy} with $u_{\cal A} = g_{\Delta {\cal A}}(\epsilon_{\cal A})$ and $u_{\cal B} = g_{\Delta {\cal B}}(\epsilon_{\cal B})$; taking again $\textsc{x} = \frac{\Delta {\cal A}}{\Delta A}$ and $\textsc{y} = \frac{\Delta {\cal B}}{\Delta B}$, this leads, after simplification and multiplication by $\Delta A \, \Delta B$, to
\ba
\frac{\Delta B + \Delta {\cal B}}{2} \ \bar \epsilon_{\cal A}^{(min)}  \ + \ \frac{\Delta A + \Delta {\cal A}}{2} \ \bar \epsilon_{\cal A}^{(min)} \, \geq \, |C_{\!AB}| \qquad \
\ea
(with again $\epsilon_{\cal A}^{(min)} = \min\big[ \epsilon_{\cal A}, \sqrt{\Delta A^2 + \Delta {\cal A}^2} \big]$ and $\epsilon_{\cal B}^{(min)} = \min\big[ \epsilon_{\cal B}, \sqrt{\Delta B^2 + \Delta {\cal B}^2} \big]$), which in turn implies Weston \emph{et al.}'s relation~\eqref{eq_weston}. In the case where $\Delta {\cal A} = 0$, Eq.~(\ref{eq:relation_g_limit_DA_0_DB_0}, left) directly implies
\ba
\frac{\Delta B{+}\Delta {\cal B}}{2} \, \epsilon_{\cal A} + \frac{\Delta A}{2} \, \epsilon_{\cal B} & \geq & \frac{\Delta B{+}\Delta {\cal B}}{2} \, \Delta A + \frac{\Delta A}{2} \, |\Delta B{-}\Delta {\cal B}| \nonumber \\
& \geq & \ \Delta A \, \Delta B \ \geq \ |C_{\!AB}|,
\ea
which gives again Weston \emph{et al.}'s relation (with $\Delta {\cal A} = 0$); the case where $\Delta {\cal B} = 0$ is obtained similarly.

Once again, the independent use of 4 CS inequalities in the proof above explains the non-optimality of Weston \emph{et al.}'s relation, as opposed to our relation~\eqref{eq:relation_g}. As before, one can check that Weston \emph{et al.}'s relation can only be saturated if $\epsilon_{\cal A} = 0$ or $\epsilon_{\cal B} = 0$, or if $\Delta {\cal A} \, \Delta {\cal B} = 0$, $\epsilon_{\cal A} = \Delta A - \Delta {\cal A}$, $\epsilon_{\cal B} = \Delta B - \Delta {\cal B}$ and $A$, $B$ and $\ket{\psi}$ saturate Robertson's relation.

Note again that one could also derive other (non-tight) inequalities from our relations of Section~\ref{sec_err_trade_off_relations}, by using Eq.~\eqref{eq:general_4cauchy} with different values of $\textsc{x}$ and $\textsc{y}$.

\section{The case of mixed states}
\label{sec_mixed_states}

So far we only considered the case where the state $\ket{\psi} \in {\cal H}$ under study is pure. In this final section we extend our analysis to mixed states and thus consider instead a density matrix $\rho \in L({\cal H})$ (where $L({\cal H})$ is the space of linear operators acting on the Hilbert space ${\cal H}$), on which the joint measurement of two (incompatible) observables $A$ and $B$ is to be approximated.

All the definitions introduced in Section~\ref{sec_defs} easily generalize; for instance, we now have $\Delta A = \Tr[(A - \moy{A})^2 \rho]^{1/2}$, $\Delta B = \Tr[(B - \moy{B})^2 \rho]^{1/2}$, $C_{\!AB} = \frac{1}{2i} \Tr \big[ [A,B] \rho \big]$ and, still assuming $\Delta A, \Delta B > 0$, $\tilde C_{\!AB} = \frac{C_{\!AB}}{\Delta A \, \Delta B}$.
Moreover, Ozawa's framework for approximate joint measurements also generalizes easily: a joint measurement of $A$ and $B$ can be approximated by the measurement of two compatible observables ${\cal A}$ and ${\cal B}$ on the system $\rho \otimes \ket{\xi}\!\bra{\xi}$ composed on the state $\rho$ and of an ancillary state $\ket{\xi}\!\bra{\xi} \in L({\cal K})$. Ozawa's inaccuracies (root-mean-square errors) $\epsilon_{\cal A}, \epsilon_{\cal B}$ are now given by
\ba
\epsilon_{\cal A}^2 &=& \Tr \big[ ({\cal A} - A)^2 \, (\rho \otimes \ket{\xi}\!\bra{\xi}) \big] , \label{eq:def_epsA_rho} \\
\epsilon_{\cal B}^2 &=& \Tr \big[ ({\cal B} - B)^2 \, (\rho \otimes \ket{\xi}\!\bra{\xi}) \big] \label{eq:def_epsB_rho}
\ea
(where again, $A$ and $B$ stand for simplicity for $A \otimes \one_{\cal K}$ and $B \otimes \one_{\cal K}$, resp.).
Note in particular that the squared inaccuracies $\epsilon_{\cal A}^2$ and $\epsilon_{\cal B}^2$ are linear in $\rho$.
The definitions of $\Delta {\cal A}, \Delta {\cal B}, \delta_{\cal A}$ and $\delta_{\cal B}$ are also trivial to generalize.

It is straightforward to check that all error-trade-off relations derived in Section~\ref{sec_err_trade_off_relations} \emph{still hold} in the case of mixed states: indeed, $\rho \in L({\cal H})$ can always be thought of as the partial trace of a pure state $\ket{\psi}$ in some extended Hilbert space ${\cal H} \otimes {\cal H'}$. By extending the observables $A$, $B$, ${\cal A}$ and ${\cal B}$ appropriately in the form $A' = A \otimes \one_{{\cal H}'}$, $B' = B \otimes \one_{{\cal H}'}$, ${\cal A}' = {\cal A} \otimes \one_{{\cal H}'}$ and ${\cal B}' = {\cal B} \otimes \one_{{\cal H}'}$, we have (with obvious notations)
\ba
& \Delta A' = \Delta A, \ \Delta B' = \Delta B, \ \tilde C_{\!A'B'} = \tilde C_{\!AB}, \nonumber \\
& \epsilon_{{\cal A}'} = \epsilon_{{\cal A}}, \ \epsilon_{{\cal B}'} = \epsilon_{{\cal B}} , \\
& \Delta {\cal A}' = \Delta {\cal A}, \ \Delta {\cal B}' = \Delta {\cal B}, \ \delta_{{\cal A}'} = \delta_{{\cal A}}, \ \delta_{{\cal B}'} = \delta_{{\cal B}} . & \qquad \nonumber
\ea
Now, the relations of Section~\ref{sec_err_trade_off_relations}, proven for pure states, necessarily hold for the primed quantities above; as the non-primed quantities are the same, the relations also hold for the latter.

However, the main difference with the case of pure states is that our relations are in general \emph{not tight} for mixed states. The reason why the proofs of Section~\ref{sec_tightness} and of the Appendix fail here is that these assumed that the approximate observables ${\cal A}$ and ${\cal B}$ could access the whole Hilbert space containing the pure state $\ket{\psi}$ (or in fact, at least the whole $\text{Span} \{ \ket{\psi}, \tilde A_0 \ket{\psi}, \tilde B_0 \ket{\psi} \}$). Now, if the mixed state $\rho$ is the partial trace of a pure state $\ket{\psi} \in {\cal H} \otimes {\cal H'}$, and one can only access and measure the state $\rho$, the possible approximate measurements ${\cal A}$ and ${\cal B}$ are restricted to act only on ${\cal H}$ (and possibly on the ancillary space ${\cal K}$), rather than on ${\cal H} \otimes {\cal H'}$. As the constructions for ${\cal A}$ and ${\cal B}$ used in Section~\ref{sec_tightness} and in the Appendix would not be restricted here to the subspace ${\cal H}$ (but would act on the space ${\cal H} \otimes {\cal H'}$ containing the purification $\ket{\psi}$), these cannot in general be used in the case of mixed states.

\subsection{A simple example}

Let us illustrate these considerations with the following example. Consider the fully mixed qubit state $\rho = \one_{\cal H}/2 = \frac{1}{2} \big( \ket{0}\!\bra{0} + \ket{1}\!\bra{1} \big)$ (with ${\cal H} = \mathbb{C}^2$), and choose the two observables $A$ and $B$ to be the Pauli operators $A = \sigma_x = \ket{1}\!\bra{0} + \ket{0}\!\bra{1}$ and $B = \sigma_y = i \ket{1}\!\bra{0} - i \ket{0}\!\bra{1}$. For this state and this pair of observables, we find (writing from now on the dependency on the state explicitly):
\ba
\Delta A (\rho) = \Delta B (\rho) = 1, \ C_{\!AB} (\rho) = 0,
\ea
and hence our error-trade-off relation~\eqref{eq:relation_PNAS} writes
\ba
\epsilon_{\cal A} (\rho)^2 + \epsilon_{\cal B} (\rho)^2 + 2 \, \epsilon_{\cal A} (\rho) \, \epsilon_{\cal B} (\rho) \ \geq \ 0 , \label{eq:rho_trivial_relation}
\ea
which is trivial and does not impose any restriction on the possible values of $(\epsilon_{\cal A}(\rho), \epsilon_{\cal B}(\rho))$.

Note however that for the states $\ket{0}$ and $\ket{1}$, one still gets $\Delta A (\ket{0}) = \Delta A (\ket{1}) = \Delta B (\ket{0}) = \Delta B (\ket{1}) = 1$, but now $C_{\!AB} (\ket{0}) = - C_{\!AB} (\ket{1}) = 1$, and our relation~\eqref{eq:relation_PNAS} imposes that
\ba
\epsilon_{\cal A} (\ket{0})^2 + \epsilon_{\cal B} (\ket{0})^2 \ \geq \ 1 , \quad \epsilon_{\cal A} (\ket{1})^2 + \epsilon_{\cal B} (\ket{1})^2 \ \geq \ 1 . \nonumber \\
\label{eq:rel_ket0_ket1}
\ea

Now, by the linearity of $\epsilon_{\cal A}^2$ and $\epsilon_{\cal B}^2$ in the density matrix, and from the decomposition $\rho = \frac{1}{2} \big( \ket{0}\!\bra{0} + \ket{1}\!\bra{1} \big)$, we have
\ba
&\epsilon_{\cal A} (\rho)^2 = \frac{1}{2} \big[ \epsilon_{\cal A} (\ket{0})^2 + \epsilon_{\cal A} (\ket{1})^2 \big], \\[1mm]
&\epsilon_{\cal B} (\rho)^2 = \frac{1}{2} \big[ \epsilon_{\cal B} (\ket{0})^2 + \epsilon_{\cal B} (\ket{1})^2 \big],
\ea
and by summing these relations we find, using~\eqref{eq:rel_ket0_ket1},
\ba
\epsilon_{\cal A} (\rho)^2 + \epsilon_{\cal B} (\rho)^2 & \geq & 1 . \label{eq:rho_nontrivial_relation}
\ea
This tells that the possible values of $(\epsilon_{\cal A}(\rho), \epsilon_{\cal B}(\rho))$ are in fact restricted\footnote{This relation strengthens significantly one given in~\cite{Buscemi:2013aa}, where it is claimed that no restrictions on $(\epsilon_{\cal A}(\rho), \epsilon_{\cal B}(\rho))$, for $\rho = \one_{\cal H}/2$, could be obtained from previously known error-trade-off relations (including~\eqref{eq:relation_PNAS}). We show here that such restrictions can in fact be obtained, indirectly.}---which our relation~\eqref{eq:relation_PNAS} failed to detect when applied directly (as we first did above in Eq.~\eqref{eq:rho_trivial_relation}). As a matter of fact, the relation~\eqref{eq:rho_nontrivial_relation} is tight: taking for instance ${\cal A} = \cos \varphi \, (\cos \varphi \, \sigma_x + \sin \varphi \, \sigma_y)$ and ${\cal B} = \sin \varphi \, (\cos \varphi \, \sigma_x + \sin \varphi \, \sigma_y)$, for any $\varphi \in \mathbb{R}$, gives $\epsilon_{\cal A} (\rho)^2 = \sin^2 \varphi$ and $\epsilon_{\cal B} (\rho)^2 = \cos^2 \varphi$, which saturate~\eqref{eq:rho_nontrivial_relation}.

It is instructive to see in this example why the tightness of the (then trivial) relation~\eqref{eq:relation_PNAS} for any purification of $\rho$ does not imply its tightness for the mixed state (partial trace) $\rho$. The fully mixed state $\rho = \one_{\cal H}/2$ can be purified to a maximally entangled state, say $\ket{\Phi^+} = \frac{1}{\sqrt{2}}(\ket{00}+\ket{11}) \in {\cal H} \otimes {\cal H}'$, with ${\cal H}' = {\cal H} = \mathbb{C}^2$. As described above, the observables $A = \sigma_x$ and $B = \sigma_y$  can be extended to $A' = \sigma_x \otimes \one_{{\cal H}'}$ and $B' = \sigma_y \otimes \one_{{\cal H}'}$, to be measured on $\ket{\Phi^+}$. According to~\eqref{eq:relation_PNAS}, and as in~\eqref{eq:rho_trivial_relation}, there is in general no restriction on $\big( \epsilon_{\cal A}(\ket{\Phi^+}), \epsilon_{\cal B}(\ket{\Phi^+}) \big)$ if the approximations have access to the extended Hilbert space; one can indeed get $\epsilon_{\cal A}(\ket{\Phi^+}) = \epsilon_{\cal B}(\ket{\Phi^+}) = 0$ by choosing for instance ${\cal A} = \ket{\Phi^+}\!\bra{\Psi^+} + \ket{\Psi^+}\!\bra{\Phi^+} + \ket{\Psi^-}\!\bra{\Psi^-}$ and ${\cal B} = i \big( \ket{\Phi^+}\!\bra{\Psi^-} - \ket{\Psi^-}\!\bra{\Phi^+} + \ket{\Psi^+}\!\bra{\Psi^-} - \ket{\Psi^-}\!\bra{\Psi^+} \big)$ (with $\ket{\Psi^\pm} = \frac{1}{\sqrt{2}}(\ket{01}\pm\ket{10})$). However, the values $\epsilon_{\cal A}(\ket{\Phi^+}) = \epsilon_{\cal B}(\ket{\Phi^+}) = 0$ cannot be obtained with any observables ${\cal A}, {\cal B}$ of the form ${\cal A}_{{\cal H} \otimes {\cal K}} \otimes \one_{{\cal H}'}, {\cal B}_{{\cal H} \otimes {\cal K}} \otimes \one_{{\cal H}'}$ (and neither can any values such that $\epsilon_{\cal A}^2 + \epsilon_{\cal B}^2 < 1$)---as this would otherwise violate the relation~\eqref{eq:rho_nontrivial_relation}.

\subsection{Strengthening our error-trade-off relations for mixed states}

The previous example shows that although our error-trade-off relations of Section~\ref{sec_err_trade_off_relations}---e.g.~\eqref{eq:relation_PNAS}---are in general not tight (and may even be trivial) for mixed states, one can still obtain stronger relations by combining them in appropriate ways.

Consider, more generally, any decomposition $\rho = \sum_i p_i \ket{\psi_i}\!\bra{\psi_i}$ (with $p_i \geq 0, \sum_i p_i = 1$) of an arbitrary mixed state $\rho$.
By linearity of $\epsilon_{\cal A}^2$ and $\epsilon_{\cal B}^2$, one has $\epsilon_{\cal A}(\rho)^2 = \sum_i p_i \, \epsilon_{\cal A}(\ket{\psi_i})^2$ and $\epsilon_{\cal B}(\rho)^2 = \sum_i p_i \, \epsilon_{\cal B}(\ket{\psi_i})^2$.
Now, the values of each pair $\big( \epsilon_{\cal A}(\ket{\psi_i}), \epsilon_{\cal B}(\ket{\psi_i}) \big)$ are restricted to satisfy in particular the (tight) relation~\eqref{eq:relation_PNAS}; let us denote the set of possible values $\big( \epsilon_{\cal A}(\ket{\psi_i}), \epsilon_{\cal B}(\ket{\psi_i}) \big)$ by ${\cal S}_{A,B}(\ket{\psi_i})$. This allows us to state the following necessary condition that the values of $\big( \epsilon_{\cal A}(\rho), \epsilon_{\cal B}(\rho) \big)$ must satisfy, for any such decomposition of $\rho$:
\ba
\hspace{-2.3cm} & \forall \, i, \ \ \exists \, (\epsilon_{\cal A}^{(i)}, \epsilon_{\cal B}^{(i)}) \in {\cal S}_{A,B}(\ket{\psi_i}), \nonumber \\
\hspace{-2.3cm} & \qquad \qquad \qquad  \quad \text{s.t. } \epsilon_{\cal A}(\rho)^2 = \sum_i p_i \, (\epsilon_{\cal A}^{(i)})^2, \ \epsilon_{\cal B}(\rho)^2 = \sum_i p_i \, (\epsilon_{\cal B}^{(i)})^2 . & \qquad \quad \
\ea
We leave open the problem of finding a systematic way to translate this constraint into a simple relation, and to systematically find the optimal decomposition that leads to a tight relation for $\big( \epsilon_{\cal A}(\rho), \epsilon_{\cal B}(\rho) \big)$---as we did above in the particular case of the 1-qubit fully mixed state.

To finish with, let us mention that the other error-trade-off relations of Section~\ref{sec_err_trade_off_relations} can also be strengthened for mixed states, using similar ideas as above. However, that may in general require one to solve even more tedious constrained optimisation problems.

\section{Conclusion}

We have derived in this paper a number of error-trade-off relations---Eqs.~\eqref{eq:relation_f}, \eqref{eq:relation_g}, \eqref{eq:relation_h}, \eqref{eq:relation_PNAS}, \eqref{eq:relation_fbis} and~\eqref{eq:relation_PRL}---bounding Ozawa's inaccuracies in the approximate joint measurement of two incompatible observables $A$ and $B$, and bounding the inaccuracy and the disturbance in a measurement-disturbance scenario. 
These relations are adapted to different cases, where one may specify certain properties of the approximations ${\cal A}$ and ${\cal B}$. They were shown to all follow from our first relation~\eqref{eq:relation_f}, which holds for some specified values of the standard deviations $\Delta {\cal A}$, $\Delta {\cal B}$, and of the biases $\delta_{\cal A}$, $\delta_{\cal B}$. As our relations are derived in the general framework of Ozawa, they could be tested with the same experimental setups as those used in Refs.~\cite{erhart2012edu,rozema2012vhm,Weston:2013fk,Sulyok:2013fk,Baek:2013ys,Ringbauer:2013aa,Kaneda:2013aa}.

We showed that our relations Eqs.~\eqref{eq:relation_f}, \eqref{eq:relation_g}, \eqref{eq:relation_h} and~\eqref{eq:relation_PNAS} are all \emph{tight} for pure states, and so are the relations~\eqref{eq:relation_fbis} and~\eqref{eq:relation_PRL} in some particular cases. In particular, they are stronger than the previously known relations of Ozawa~\cite{ozawa2003uvr,ozawa2004urj}, Hall~\cite{hall2004pih} and Weston \emph{et al.}~\cite{Weston:2013fk}, which could be explicitly derived from our relations directly. Our study has allowed us to clarify the difference between the similar-looking relations of Ozawa and Hall, as well as with Weston \emph{et al.}'s relation, and to pinpoint precisely why their derivations are non-optimal. The question of the tightness of our relations for mixed states has also been addressed; we have shown how to combine our (no longer tight) relations and still obtain possibly tight constraints for mixed states.
The question of the tightness of error-trade-off or error-disturbance relations is quite relevant indeed: tight relations give not only negative results (by quantifying ``what \emph{cannot} be done quantum mechanically''), but also positive results (by specifying ``what \emph{can} be done''). In our proofs that our relations are tight, we constructed explicit measurement schemes allowing one to obtain any values of $(\epsilon_{\cal A}, \epsilon_{\cal B})$ that saturate our relations.

Our work has been focused on Ozawa's framework and definitions for measurement inaccuracies.
As mentioned in the Introduction however, other approaches have been (and certainly will be) proposed to quantify the ``measurement uncertainty'' aspect of Heisenberg's Uncertainty Principle. In the prospect of possible applications in quantum information science, a promising direction of research is for instance the study of \emph{entropic} relations, as recently considered in Refs.~\cite{Buscemi:2013aa,Coles:2013aa}. It will be interesting to see if any of the techniques used in this paper (e.g. the cascaded derivation of error-trade-off relations from one another, the study of their tightness) could be adapted and used in other contexts, where another approach to quantifying Heisenberg's Uncertainty Principle is studied.

\section*{Acknowledgments}

This work was supported by a University of Queensland Postdoctoral Research Fellowship.

\appendix

\section*{Appendix: Tightness of our error-trade-off relations~(\ref{eq:relation_f}), (\ref{eq:relation_g}), (\ref{eq:relation_h}) and (\ref{eq:relation_PNAS}) for $\big| \moy{\tilde A_0 \tilde B_0} \big| < 1$}

In this Appendix we prove that our relations~\eqref{eq:relation_f}, \eqref{eq:relation_g}, \eqref{eq:relation_h} and \eqref{eq:relation_PNAS} are tight when $\big| \moy{\tilde A_0 \tilde B_0} \big| < 1$, thus completing the proof of Section~\ref{sec_tightness} in which the case $\big| \moy{\tilde A_0 \tilde B_0} \big| = 1$ was considered.
The proof here follows very similar lines to that of the main text, and just involves slightly more tedious calculations.
For simplicity we shall not repeat all the details of the proof, but insist on what is different from Section~\ref{sec_tightness}.

\medskip

Let us again define $\phi = \arg \, \moy{\tilde A_0 \tilde B_0} \in [-\pi,\pi]$, and let us now also introduce $r = \big| \moy{\tilde A_0 \tilde B_0} \big|$, with $0 \leq r < 1$ in the case considered here, and $\phi' = \arcsin \tilde C_{\!AB} \in$ \mbox{$]{-}\frac{\pi}{2}, \frac{\pi}{2}[$}---such that $\moy{\tilde A_0 \tilde B_0} = r e^{i \phi}$ and $\tilde C_{\!AB} = {\mathrm{Im}} \moy{\tilde A_0 \tilde B_0} = r \, \sin \phi = \sin \phi'$.
The values of $(u_{\cal A}, u_{\cal B})$ that saturate the general relation~\eqref{eq:relation_general_form} (or~\eqref{eq:relation_general_form_altern}, equivalently) can then be parametrized by
\ba
\Big( \, u_{\cal A} = \Big| \sin \Big( \frac{\varphi + \phi'}{2} \Big) \Big|, \ u_{\cal B} = \Big| \sin \Big( \frac{\varphi - \phi'}{2} \Big) \Big| \ \Big) , \quad \ \ \label{eq_param_ua_ub_saturation_d3}
\ea
for a varying value of $\varphi \in \big[ {-}|\phi'|, |\phi'| \big]$.
As in Section~\ref{sec_tightness}, in order to show the tightness of our error-trade-off relations, we will show that the corresponding functions $(u_{\cal A}, u_{\cal B})$ can indeed take these values.

It will be convenient below to use the notations $c_{\phi^{(\prime)}} = \cos \phi^{(\prime)}$ and $s_{\phi^{(\prime)}} = \sin \phi^{(\prime)}$ (such that $r \, s_\phi = s_{\phi'}$), and to define $q = \frac{c_\phi}{c_{\phi'}}$ (note that $c_{\phi'} \geq |c_\phi|$ and $c_{\phi'} > 0$, so that $|q| \leq 1$).

\subsection{Parametrization of ${\cal A}$ and ${\cal B}$}

Let us first introduce a different parametrization\footnote{Note that the parametrization introduced in Section~\ref{sec_tightness} cannot be used in the case $\big| \moy{\tilde A_0 \tilde B_0} \big| < 1$, as the unit vectors $\tilde A_0 \ket{\psi}$ and $\tilde B_0 \ket{\psi}$ are no longer the same (up to a phase). Our parametrization here is again somewhat clearer, but equivalent to that used in~\cite{branciard2013ete}.} for the approximate observables ${\cal A}$ and ${\cal B}$, acting on (the now 3-dimensional) $\text{Span} \{ \ket{\psi}, \tilde A_0 \ket{\psi}, \tilde B_0 \ket{\psi} \}$.

We first define
\ba
\ket{v_1} &=& \ket{\psi}, \\
\ket{v_2} &=& \frac{e^{i \phi/2} \tilde A_0 + e^{-i \phi/2} \tilde B_0}{\sqrt{2+2r}} \ket{\psi}, \\
\ket{v_3} &=& \frac{e^{i \phi/2} \tilde A_0 - e^{-i \phi/2} \tilde B_0}{\sqrt{2-2r}} \ket{\psi},
\ea
so that $\{ \ket{v_1}, \ket{v_2}, \ket{v_3} \}$ forms an orthonormal basis of $\text{Span} \{ \ket{\psi}, \tilde A_0 \ket{\psi}, \tilde B_0 \ket{\psi} \}$.
Note that $\tilde A_0 \ket{\psi}$ and $\tilde B_0 \ket{\psi}$ are then given by
\ba
\tilde A_0 \ket{\psi} & = & e^{-i \phi/2} \Big( \sqrt{\frac{1+r}{2}} \ket{v_2} + \sqrt{\frac{1-r}{2}} \ket{v_3} \Big) , \quad \\
\tilde B_0 \ket{\psi} & = & e^{i \phi/2} \Big( \sqrt{\frac{1+r}{2}} \ket{v_2} - \sqrt{\frac{1-r}{2}} \ket{v_3} \Big) .
\ea

For a given $\varphi \in \big[ {-}|\phi'|, |\phi'| \big]$, and for any 3 parameters $\theta_1, \theta_2, \theta_3 \in \mathbb{R}$ such that $\cos \theta_1 \sin \theta_1 \cos \theta_2 \sin \theta_2 \neq 0$, let us define (with $c_j = \cos \theta_j$ and $s_j = \sin \theta_j$ for $j = 1,2,3$) the complex parameters
\ba
\gamma_1 & = & \sqrt{\frac{1{+}q}{2}} \, c_3 \, e^{i \varphi/2}  - i \, \text{sign}(s_\phi) \, \sqrt{\frac{1{-}q}{2}} \, s_3 \, e^{-i \varphi/2} , \qquad \ \\
\gamma_2 & = & \sqrt{\frac{1{+}q}{2}} \, s_3 \, e^{-i \varphi/2}  - i \, \text{sign}(s_\phi) \, \sqrt{\frac{1{-}q}{2}} \, c_3 \, e^{i \varphi/2} ,
\ea
such that $|\gamma_1|^2 + |\gamma_2|^2 = 1$, and
\ba
\ket{m_1} &=& c_1 \ket{v_1} -s_1 \gamma_1 \ket{v_2} -s_1 \gamma_2 \ket{v_3}, \nonumber \\[1mm]
\ket{m_2} &=& s_1 c_2 \ket{v_1} + (c_1 c_2 \gamma_1{+}s_2 \gamma_2^*) \ket{v_2} + (c_1 c_2 \gamma_2{-}s_2 \gamma_1^*) \ket{v_3}, \nonumber \\[1mm]
\ket{m_3} &=& s_1 s_2 \ket{v_1} + (c_1 s_2 \gamma_1{-}c_2 \gamma_2^*) \ket{v_2} + (c_1 s_2 \gamma_2{+}c_2 \gamma_1^*) \ket{v_3}, \nonumber \\
\ea
(where $\gamma_j^*$ denotes the complex conjugate of $\gamma_j$),
so that $\{ \ket{m_1}, \ket{m_2}, \ket{m_3} \}$ also forms an orthonormal basis of $\text{Span} \{ \ket{\psi}, \tilde A_0 \ket{\psi}, \tilde B_0 \ket{\psi} \}$. If the dimension of ${\cal H}$ is larger than 3, we complete that basis with other vectors $\ket{m_{k \geq 4}}$ orthogonal to $\ket{m_1}$, $\ket{m_2}$ and $\ket{m_3}$ (which are hence orthogonal to $\ket{\psi}$).

As in Section~\ref{sec_tightness}, the basis $\{ \ket{m_k} \}$ is chosen to be the common eigenbasis of ${\cal A}$ and ${\cal B}$, defined again through Eq.~\eqref{eq_def_Aest_Best}. With this parametrization, we get
\ba
\moy{{\cal A}} &=& \alpha_1 \, c_1^2 + \alpha_2 \, s_1^2 \, c_2^2 + \alpha_3 \, s_1^2 \, s_2^2, \qquad \delta_{\cal A} = \moy{{\cal A}} - \moy{A}, \nonumber \\[1mm]
\Delta {\cal A}^2 &=& (\alpha_1{-}\moy{{\cal A}})^2 c_1^2 + (\alpha_2{-}\moy{{\cal A}})^2 s_1^2 c_2^2 + (\alpha_3{-}\moy{{\cal A}})^2 s_1^2 s_2^2, \nonumber \\ \label{eq_dAest_DAest_d3} \\[2mm]
\moy{{\cal B}} &=& \beta_1 \, c_1^2 + \beta_2 \, s_1^2 \, c_2^2 + \beta_3 \, s_1^2 \, s_2^2, \qquad \delta_{\cal B} = \moy{{\cal B}} - \moy{B}, \nonumber \\[1mm]
\Delta {\cal B}^2 &=& (\beta_1{-}\moy{{\cal B}})^2 c_1^2 + (\beta_2{-}\moy{{\cal B}})^2 s_1^2 c_2^2 + (\beta_3{-}\moy{{\cal B}})^2 s_1^2 s_2^2. \nonumber \\ \label{eq_dBest_DBest_d3}
\ea

\subsection{Calculation of $\epsilon_{\cal A}$ and $\epsilon_{\cal B}$}

Defining
\ba
F_\pm &=& \sqrt{\frac{1+q}{2}} \sqrt{\frac{1 \pm r}{2}} \cos \Big( \frac{\varphi \pm \phi}{2} \Big) \nonumber \\ 
&& \ \pm \ \text{sign}(s_\phi) \sqrt{\frac{1-q}{2}} \sqrt{\frac{1 \mp r}{2}} \sin \Big( \frac{\varphi \pm \phi}{2} \Big) , \qquad \\
G_\pm &=& \pm \sqrt{\frac{1+q}{2}} \sqrt{\frac{1 \pm r}{2}} \cos \Big( \frac{\varphi \mp \phi}{2} \Big) \nonumber \\ 
&& \ - \ \text{sign}(s_\phi) \sqrt{\frac{1-q}{2}} \sqrt{\frac{1 \mp r}{2}} \sin \Big( \frac{\varphi \mp \phi}{2} \Big) , \qquad
\ea
we find, after some calculations,
\ba
{\mathrm{Re}} \, \sandwich{\psi}{\tilde A_0}{m_1} \braket{m_1}{\psi} &=& - c_1 s_1 (c_3 F_+{+}s_3 F_-) , \label{eq_Re_psi_A0_m1_m1_psi_d3} \\
{\mathrm{Re}} \, \sandwich{\psi}{\tilde A_0}{m_2} \braket{m_2}{\psi} &=& s_1 c_2 \big[ c_1 c_2 (c_3 F_+{+}s_3 F_-) \nonumber \\ && \qquad \ + s_2 (s_3 F_+{-}c_3 F_-) \big] , \qquad \label{eq_Re_psi_A0_m2_m2_psi_d3} \\
{\mathrm{Re}} \, \sandwich{\psi}{\tilde A_0}{m_3} \braket{m_3}{\psi} &=& s_1 s_2 \big[ c_1 s_2 (c_3 F_+{+}s_3 F_-) \nonumber \\ && \qquad \ - c_2 (s_3 F_+{-}c_3 F_-) \big] , \qquad \label{eq_Re_psi_A0_m3_m3_psi_d3}
\ea
and similarly for ${\mathrm{Re}} \, \sandwich{\psi}{\tilde B_0}{m_k} \braket{m_k}{\psi}$, where $F_\pm$ are replaced by $G_\pm$.

Note further that
\ba
(c_3 F_+{+}s_3 F_-)^2 + (s_3 F_+{-}c_3 F_-)^2 \ = \ F_+^2 + F_-^2 \hspace{-7cm} \nonumber \\
&=& \frac{1}{2} \Big( 1 + \big( q \, c_\phi \!+\! \sqrt{1{-}q^2} \sqrt{1{-}r^2} \, |s_\phi| \big) \cos \varphi - r s_\phi \, \sin \varphi \Big) \nonumber \\
&=& \frac{1}{2} \Big( 1 + \Big( \frac{c_\phi^2}{c_{\phi'}} + \sqrt{\frac{c_{\phi'}^2{-}c_\phi^2}{c_{\phi'}^2}} \sqrt{s_\phi^2{-}s_{\phi'}^2} \, \Big) \cos \varphi - s_{\phi'} \, \sin \varphi \Big) \nonumber \\
&=& \frac{1}{2} \Big( 1 + c_{\phi'} \cos \varphi - s_{\phi'} \, \sin \varphi \Big) = \cos^2 \Big( \frac{\varphi + \phi'}{2} \Big),
\ea
and similarly,
\ba
(c_3 G_+{+}s_3 G_-)^2 + (s_3 G_+{-}c_3 G_-)^2 = \cos^2 \! \Big( \! \frac{\varphi - \phi'}{2} \! \Big) . \quad \quad \ \
\ea
This implies that there exist two real parameters $\theta_a, \theta_b$ such that, with $c_a = \cos \theta_a, s_a = \sin \theta_a, c_b = \cos \theta_b$ and $s_b = \sin \theta_b$,
\ba
c_3 F_+ + s_3 F_- &=& c_a \, \cos \Big( \frac{\varphi + \phi'}{2} \Big), \label{eq_c3Fp_s3Fm} \\
s_3 F_+ - c_3 F_- &=& s_a \, \cos \Big( \frac{\varphi + \phi'}{2} \Big), \label{eq_s3Fp_c3Fm} \\[2mm]
c_3 G_+ + s_3 G_- &=& c_b \, \cos \Big( \frac{\varphi - \phi'}{2} \Big), \label{eq_c3Gp_s3Gm} \\
s_3 G_+ - c_3 G_- &=& s_b \, \cos \Big( \frac{\varphi - \phi'}{2} \Big). \label{eq_s3Gp_c3Gm}
\ea
Eqs.~(\ref{eq:relation_eps_delta_Deltas_A}--\ref{eq:relation_eps_delta_Deltas_B}) (with ${\mathrm{Re}} \, \moy{\tilde A_0 {\cal A}} = \sum \alpha_k \, {\mathrm{Re}} \, \sandwich{\psi}{\tilde A_0}{m_k}\braket{m_k}{\psi}$ and ${\mathrm{Re}} \, \moy{\tilde B_0 {\cal B}} = \sum \beta_k \, {\mathrm{Re}} \, \sandwich{\psi}{\tilde B_0}{m_k}\braket{m_k}{\psi}$), together with Eqs.~(\ref{eq_Re_psi_A0_m1_m1_psi_d3}--\ref{eq_Re_psi_A0_m3_m3_psi_d3}), then imply
\ba
\epsilon_{\cal A}^2 - \delta_{\cal A}^2 &=& \Delta A^2 + \Delta {\cal A}^2 - 2 \, \Delta A \ X_\alpha \, \cos \Big( \frac{\varphi + \phi'}{2} \Big), \qquad \ \label{eq:epsA_explicit_d3} \\
\epsilon_{\cal B}^2 - \delta_{\cal B}^2 &=& \Delta B^2 + \Delta {\cal B}^2 - 2 \, \Delta B \ Y_\beta \, \cos \Big( \frac{\varphi - \phi'}{2} \Big). \qquad \label{eq:epsB_explicit_d3}
\ea
with
\ba
X_\alpha &=& s_1 \big[ ({-}\alpha_1 {+} \alpha_2 \, c_2^2 {+} \alpha_3 \, s_2^2 ) \, c_1 c_a + (\alpha_2{-}\alpha_3) \, c_2 s_2 s_a \big] , \nonumber \\[1mm]
Y_\beta &=& s_1 \big[ ({-}\beta_1 {+} \beta_2 \, c_2^2 {+} \beta_3 \, s_2^2 ) \, c_1 c_b + (\beta_2{-}\beta_3) \, c_2 s_2 s_b \big] . \nonumber \\
\ea

\subsection{Tightness of relation~\eqref{eq:relation_f}}

We shall now only sketch the calculations that allow us to prove the tightness of our relations~\eqref{eq:relation_f}, \eqref{eq:relation_g}, \eqref{eq:relation_h} and \eqref{eq:relation_PNAS}, and refer to Section~\ref{sec_tightness} of the main text for more detailed explanations.

\medskip

Relation~\eqref{eq:relation_f} holds for some specified values of $\Delta {\cal A}, \Delta {\cal B}, \delta_{\cal A}$ and $\delta_{\cal B}$. For these values and for some choice of $\tau_\alpha, \tau_\beta = \pm 1$, we define the eigenvalues of ${\cal A}$ and ${\cal B}$ to be
\ba
\alpha_1 &=& \moy{A} + \delta_{{\cal A}} \, - \tau_\alpha \, \frac{s_1 c_a}{c_1} \, \Delta {\cal A} , \label{eq_alpha1_for_f_d3} \\
\alpha_2 &=& \moy{A} + \delta_{{\cal A}} \, + \tau_\alpha \, \frac{c_1 c_2 c_a + s_2 s_a}{s_1 c_2} \, \Delta {\cal A} , \label{eq_alpha2_for_f_d3} \\
\alpha_3 &=& \moy{A} + \delta_{{\cal A}} \, + \tau_\alpha \, \frac{c_1 s_2 c_a - c_2 s_a}{s_1 s_2} \, \Delta {\cal A} , \quad \label{eq_alpha3_for_f_d3} \\[2mm]
\beta_1 &=& \moy{B} + \delta_{{\cal B}} \, - \tau_\beta \, \frac{s_1 c_b}{c_1} \, \Delta {\cal B} , \label{eq_beta1_for_f_d3} \\
\beta_2 &=& \moy{B} + \delta_{{\cal B}} \, + \tau_\beta \, \frac{c_1 c_2 c_b + s_2 s_b}{s_1 c_2} \, \Delta {\cal B} ,\label{eq_beta2_for_f_d3} \\
\beta_3 &=& \moy{B} + \delta_{{\cal B}} \, + \tau_\beta \, \frac{c_1 s_2 c_b - c_2 s_b}{s_1 s_2} \, \Delta {\cal B} . \quad\label{eq_beta3_for_f_d3}
\ea
Note that these are defined such that ${\cal A}$ and ${\cal B}$ indeed give the desired values of $\Delta {\cal A}, \Delta {\cal B}, \delta_{\cal A}, \delta_{\cal B}$ (see~(\ref{eq_dAest_DAest_d3}--\ref{eq_dBest_DBest_d3})), and furthermore satisfy
\ba
X_\alpha = \tau_\alpha \, \Delta {\cal A} , \quad
Y_\beta = \tau_\beta \, \Delta {\cal B} . \label{eq_XA_YB}
\ea

Eqs.~(\ref{eq:epsA_explicit_d3}--\ref{eq:epsB_explicit_d3}) then imply that Eqs.~(\ref{eq_frac_A}--\ref{eq_frac_B}) hold again for $\Delta {\cal A}, \Delta {\cal B} > 0$ (with $\phi$ replaced by $\phi'$, and independently of $\theta_1, \theta_2, \theta_3$), which, as in Subsection~\ref{subsec_tightness_f}, leads to the conclusion that relation~\eqref{eq:relation_f} is tight.
In the case where $\Delta {\cal A} = 0$ or $\Delta {\cal B} = 0$, the above choice of eigenvalues~(\ref{eq_alpha1_for_f_d3}--\ref{eq_beta3_for_f_d3}) also allows one to saturate the constraints of Eq.~\eqref{eq:relation_f_limit_DA_0} or~\eqref{eq:relation_f_limit_DB_0}.

\subsection{Tightness of relation~\eqref{eq:relation_g}}

In order to prove the tightness of relation~\eqref{eq:relation_g}, which holds for specified values of $\Delta {\cal A}$ and $\Delta {\cal B}$, one can as in Subsection~\ref{subsec_tightness_g} simply set $\delta_{{\cal A}} = \delta_{{\cal B}} = 0$ and $\tau_\alpha = \text{sign} [\cos \big( \frac{\varphi + \phi'}{2} \big)]$, $\tau_\beta = \text{sign} [\cos \big( \frac{\varphi - \phi'}{2} \big)]$ in the previous definitions~(\ref{eq_alpha1_for_f_d3}--\ref{eq_beta3_for_f_d3}).

These still give the desired values for $\Delta {\cal A}$ and $\Delta {\cal B}$, and lead now, for $\Delta {\cal A}, \Delta {\cal B} > 0$, to Eqs.~(\ref{eq_frac_A_g}--\ref{eq_frac_B_g}) (with $\phi \to \phi'$)---which, as in Subsection~\ref{subsec_tightness_g}, leads to the conclusion that relation~\eqref{eq:relation_g} is tight.
In the case where $\Delta {\cal A} = 0$ or $\Delta {\cal B} = 0$, the given choice of eigenvalues also allows one to saturate the constraints of Eq.~\eqref{eq:relation_g_limit_DA_0_DB_0}.

\subsection{Tightness of relation~\eqref{eq:relation_h}}

To prove the tightness of relation~\eqref{eq:relation_h}, which holds for specified values of $\delta_{\cal A}$ and $\delta_{\cal B}$, one can now, as in Subsection~\ref{subsec_tightness_h}, set $\Delta {\cal A} = \Delta A |\cos \big( \frac{\varphi + \phi'}{2} \big)|$, $\Delta {\cal B} = \Delta B |\cos \big( \frac{\varphi - \phi'}{2} \big)|$ and $\tau_\alpha = \text{sign} [\cos \big( \frac{\varphi + \phi'}{2} \big)]$, $\tau_\beta = \text{sign} [\cos \big( \frac{\varphi - \phi'}{2} \big)]$ in the definitions~(\ref{eq_alpha1_for_f_d3}--\ref{eq_beta3_for_f_d3}).

With these definitions, ${\cal A}$ and ${\cal B}$ indeed give the desired values of $\delta_{\cal A}$ and $\delta_{\cal B}$, and we get
\ba
X_\alpha = \Delta A \, \cos \Big( \frac{\varphi{+}\phi'}{2} \Big), \
Y_\beta = \Delta B \, \cos \Big( \frac{\varphi{-}\phi'}{2} \Big), \qquad
\ea
so that Eqs.~(\ref{eq:epsA_explicit_d3}--\ref{eq:epsB_explicit_d3}) imply that Eqs.~(\ref{eq_e2_d2_A}--\ref{eq_e2_d2_B}) hold again (with $\phi \to \phi'$)---which, as in Subsection~\ref{subsec_tightness_h}, leads to the conclusion that relation~\eqref{eq:relation_h} is tight.

\subsection{Tightness of relation~\eqref{eq:relation_PNAS}}

To finally prove the tightness of relation~\eqref{eq:relation_PNAS}, one can set $\delta_{{\cal A}} = \delta_{{\cal B}} = 0$, $\Delta {\cal A} = \Delta A |\cos \big( \frac{\varphi + \phi'}{2} \big)|$, $\Delta {\cal B} = \Delta B |\cos \big( \frac{\varphi - \phi'}{2} \big)|$, and $\tau_\alpha = \text{sign} [\cos \big( \frac{\varphi + \phi'}{2} \big)]$, $\tau_\beta = \text{sign} [\cos \big( \frac{\varphi - \phi'}{2} \big)]$ in the choice of eigenvalues~(\ref{eq_alpha1_for_f_d3}--\ref{eq_beta3_for_f_d3}). Note that this amounts to defining $\alpha_k = {\mathrm{Re}} \frac{\sandwich{m_k}{A}{\psi}}{\braket{m_k}{\psi}}$ and $\beta_k = {\mathrm{Re}} \frac{\sandwich{m_k}{B}{\psi}}{\braket{m_k}{\psi}}$, as already noted in Subsection~\ref{subsec_tightness_h}. We find that~\eqref{eq_epsA_epsB_sinP_sinM} still holds (with $\phi \to \phi'$), which proves the tightness of relation~\eqref{eq:relation_PNAS}.

\end{document}